\documentclass[aps,prd,10pt,nofootinbib,twocolumn,superscriptaddress,preprintnumbers,balancelastpage,longbibliography]{revtex4-2}

\PassOptionsToPackage{dvipsnames}{xcolor}
\usepackage{orcidlink}
\usepackage{aas_macros}
\usepackage{appendix}
\usepackage[utf8]{inputenc}
\usepackage{amsmath,amssymb,mathtools,bm}
\usepackage{graphicx, hepunits}
\usepackage{xcolor}
\usepackage{multirow}
\usepackage{hyperref}
\usepackage{fontawesome}

\usepackage{array}
\usepackage{booktabs}
\usepackage[normalem]{ulem}

\makeatletter
\def\hlinewd#1{%
\noalign{\ifnum0=`}\fi\hrule \@height #1 \futurelet
\reserved@a\@xhline}
\makeatother

\newcolumntype{L}[1]{>{\raggedright\let\newline\\\arraybackslash\hspace{0pt}}m{#1}}
\newcolumntype{C}[1]{>{\centering\let\newline\\\arraybackslash\hspace{0pt}}m{#1}}
\newcolumntype{R}[1]{>{\raggedleft\let\newline\\\arraybackslash\hspace{0pt}}m{#1}}

\hypersetup{
    colorlinks=true,       
    linkcolor=blue,        
    citecolor=blue,        
    filecolor=magenta,     
    urlcolor=blue          
}
\usepackage[english]{babel}
\let\vec\mathbf
\usepackage{tensor}

\newcommand{\m}{m_a}
\newcommand{\g}{g_{a\gamma\gamma}}

\begin{document}

\title{Re-examining the sensitivity of JWST to decaying axion dark matter}

\author{Caleb Gemmell\,\orcidlink{0000-0002-6505-8559}}
\email{cgemmell2@wisc.edu}
\affiliation{Department of Physics, University of Wisconsin-Madison, Madison, WI 53706, U.S.A.}

\author{Christopher Dessert\,\orcidlink{0000-0003-1994-088X}}
\email{Christopher.Dessert@tufts.edu}
\affiliation{Institute of Cosmology, Department of Physics and Astronomy, Tufts University, Medford, MA 02155, U.S.A.}
\affiliation{Center for Computational Astrophysics, Flatiron Institute, New York, NY 10010, U.S.A.}

\author{Andrea Caputo\,\orcidlink{0000-0003-1122-6606}}
\affiliation{Department of Theoretical Physics, CERN, Esplanade des Particules 1, P.O. Box 1211, Geneva 23, Switzerland}
\affiliation{Dipartimento di Fisica, ``Sapienza'' Universit\`a di Roma \& Sezione INFN Roma, Piazzale Aldo Moro 5, 00185, Roma, Italy}
\affiliation{Department of Particle Physics and Astrophysics,
Weizmann Institute of Science, Rehovot 7610001, Israel}

\author{Joshua W. Foster\,\orcidlink{0000-0002-7399-2608}}
\affiliation{Department of Physics, University of Wisconsin-Madison, Madison, WI 53706, U.S.A.}

\date{\today}

\begin{abstract}
An eV-scale QCD axion comprising the observed dark matter (DM) abundance is expected to generate a photon line at infrared energies that would be observable or near-observable in data collected by the James Webb Space Telescope (JWST), as might more general axion-like particles (ALPs) over a broader range of masses and couplings. This has motivated a number of efforts to either forecast JWST sensitivities to a QCD axion or realize them through analyses of publicly available datasets. At present, no consensus exists; leading analyses disagree by as much as an order of magnitude in terms of axion-coupling sensitivity, implying an orders-of-magnitude discrepancy in raw flux density sensitivity, and consistency between the analyses and prior forecasts is unclear. We address these outstanding discrepancies with a bespoke data reduction and flexible nonparametric inference procedure that lead to well-controlled and robust limits on the decay of eV-scale axion DM, consistent with previously forecasted sensitivities. We further demonstrate that the strongest previously claimed sensitivities exceed those attainable by any analysis of the datasets from which they were derived. We exclude QCD axion DM for masses between $500\,\mathrm{meV}$ and $2.5\,\mathrm{eV}$ using NIRSpec data, while setting limits on ALP DM complementary to other astrophysical constraints at masses between $100\,\mathrm{meV}$ and $500\,\mathrm{meV}$. However, we find the sensitivities to be systematically limited, and therefore unlikely to be improved upon by ongoing data collection or re-analysis unless instrumental modeling and data reduction pipelines improve considerably.
\end{abstract}
\maketitle

\section{Introduction}

Monochromatic features at energies unassociated with Standard Model (SM) thresholds have long been identified as potential smoking-gun signatures of dark matter (DM) decay or annihilation~\cite{PhysRevLett.56.263,PhysRevD.37.3737}. Searches for $\gamma$-ray lines from GeV-scale DM are particularly clean, as no astrophysical process is known to produce narrow features at GeV energies~\cite{Bringmann:2012ez}, and $\gamma$-ray line searches provide strong bounds on DM annihilation~\cite{Foster:2022nva} even as the photon final-state is typically loop-suppressed~\cite{Bergstrom:1997fh}. In the X-ray, searches are also relatively clean, although atomic transitions in hot plasmas lead to a sparse set of emission line backgrounds~\cite{Abazajian:2001vt}. Moreover, at these energies telescopes operate as photon counters, so that the statistical uncertainty on measured fluxes is Poissonian and does not require external calibration.

On the other hand, the indirect detection of eV-scale DM requires observations in the infrared, where the sky is much brighter and contains a dense forest of narrow emission lines arising from atomic and molecular transitions, making it difficult to pick out a putative DM line. Furthermore, infrared detectors operate as integrating arrays rather than photon counters, so that computing flux uncertainties relies on input from extensive on-the-ground and in-flight calibration. A line search in this band is then limited not only by photon statistics but also by systematics associated with confounding astrophysical emission lines and the degree to which the reported uncertainties reflect the combination of statistical and systematic variation realized in the data.

Despite the associated challenges, the infrared band is a promising energy range in which to look for DM decays; in particular, it hosts expected signatures of the quantum chromodynamics (QCD) axion~\cite{Peccei:1977hh, Weinberg:1977ma, Wilczek:1977pj}, among the most well-motivated candidates for physics beyond the SM. If realized in the spectrum of nature, the QCD axion would resolve the Strong \textit{CP} Problem, and if realized in abundance, could account for the present-day cosmological abundance of DM~\cite{Preskill:1982cy, Abbott:1982af, Dine:1982ah, Cirelli:2024ssz}. Considerable effort has been devoted toward searching for axions with a mass between $1$--$100\,\mu\mathrm{eV}$~\cite{Adams:2022pbo}, but QCD axions which solve the Strong \textit{CP} problem can take masses as large as $\sim 1\,\mathrm{eV}$~\cite{DiLuzio:2016sbl, DiLuzio:2020wdo} while still realizing the correct DM abundance in a range of viable cosmologies~\cite{OHare:2024nmr, Takahashi:2023vhv, Cyncynates:2023esj}. At these relatively large masses, QCD axions, through their coupling to electromagnetism, would generate potentially observable spectral lines when they decay to photons. Axion-like particles (ALPs)~\cite{Arvanitaki:2009fg, Svrcek:2006yi}, such as those realized in string theory, could also account for the observed DM abundance and would generate similar photonic signatures. 

Searches for the decay of axion DM to optical and infrared photons have a long history~\cite{Grin:2006aw, Kohri:2017ljt, Caputo:2020msf, Nakayama:2022jza, Carenza:2023qxh, Bessho:2022yyu, Wang:2023imi}, with related searches forecast for upcoming wide-field missions~\cite{Regis:2024znx, Libanore:2024hmq}. A compelling possibility, first identified in \cite{Janish:2023kvi, Roy:2023omw}, was that the James Webb Space Telescope (JWST)~\cite{Gardner:2006ky}, when operating in spectroscopic observing modes, could have sensitivity to axion DM which decays at the rate predicted by the generic coupling of the QCD axion to electromagnetism. Efforts to realize these projected constraints with analysis of publicly available JWST data were taken in \cite{Pinetti:2025owq, Saha:2025any}, but with somewhat discrepant results. While the anticipated sensitivities of \cite{Janish:2023kvi} and \cite{Roy:2023omw} were roughly in agreement, the analysis of \cite{Pinetti:2025owq} resulted in stronger constraints than projected, while the analysis of \cite{Saha:2025any} yielded compatible but slightly weaker ones. In total, though \cite{Pinetti:2025owq, Saha:2025any} appeared nearly contemporaneously and drew on the same public JWST archive, \cite{Pinetti:2025owq} realized constraints nearly 10 times stronger in axion-coupling than \cite{Saha:2025any}, corresponding to flux density limits on infrared (IR) spectral lines that differ by a factor of 100. As a result, though both \cite{Pinetti:2025owq} and \cite{Saha:2025any} report upper limits from their search for an axion-induced line, \cite{Saha:2025any} rules out only QCD axion DM with mass $m_a \gtrsim 1\,\mathrm{eV}$ while \cite{Pinetti:2025owq} more optimistically rules out QCD axion  DM with mass $m_a \gtrsim 0.1 \,\mathrm{eV}$.

In this work, we endeavor to determine precisely the sensitivity of JWST to decaying axion DM through a re-analysis of the public data now available nearly four years after first light. This work is timely as the differing sensitivities of \cite{Pinetti:2025owq, Saha:2025any} have remained unexplained, but is also generally motivated as the now mature state of JWST data taking and data reduction mean that the presently attainable sensitivities, and the degree to which sensitivity will continue to accumulate over the anticipated additional fifteen years of operation, can now be assessed. More generally, a central focus of this work is to determine the sensitivity of infrared detectors to DM decays by developing search strategies aware of and robust to the myriad challenges in this band.

Our detailed investigation also interrogates the origin of the discrepancy between \cite{Pinetti:2025owq} and \cite{Saha:2025any}. We find generally good agreement with Ref.~\cite{Saha:2025any} at the level of individual datasets. Their reported limits, however, derive from the single most constraining dataset at each mass, and so do not make maximum use of the available data. Our joint analysis does, though its expected sensitivities are only very slightly stronger, and after our correction for systematic mismodeling, which has no analog in their analysis, our final limits are weaker than theirs by roughly $30$--$50\%$. By contrast, we find the considerably stronger sensitivities claimed in \cite{Pinetti:2025owq} to be non-reproducible. A direct re-analysis of the GN-z11 observations studied in the predecessor work \cite{Janish:2023kvi} reveals several minor errors in the associated public analysis code but broadly consistent results, providing an independent cross-check of our full analysis pipeline at the level of an individual dataset. So validated, our Fisher-information projections demonstrate that the sensitivities claimed in \cite{Pinetti:2025owq} exceed those attainable by any analysis of the datasets on which they are based, once ten products with demonstrably unphysical reported uncertainties are set aside. Separately, we find the archival PRISM observations that our fiducial analysis excludes, though nominally powerful, to also be highly systematically limited, and the limits they yield after our mismodeling correction are comparatively less constraining.

Trustworthy limits also require that the ensemble of discovery significances across the searched masses be consistent with the null expectation. We test this consistency directly, finding gross inconsistency with the null prior to our data-driven mismodeling correction, which restores it by construction. That known astrophysical lines remain significant after this correction demonstrates that it preserves sensitivity to genuine line-like signals, while signal injection tests verify the recovery of injected signals. The degree of mismodeling also determines how sensitivity scales with data volume. As observations are added to the joint analysis, we find the expected statistical sensitivity improves at the anticipated rate, but the significance of the mismodeling grows along with it, requiring data-driven correction. We then find that the robust sensitivity saturates, or even weakens, with increasing data volume, and therefore do not anticipate that continued data taking or re-analysis will meaningfully improve upon the sensitivities reported here unless instrumental modeling and data-reduction pipelines improve considerably.

Our work is structured as follows. In Sec.~\ref{sec:IR_Lines_Review}, we review how axions or other decaying DM candidates generate IR lines through decay to photons; we also develop a detailed modeling pipeline that accounts for the attenuation of DM-induced lines through the absorption and scattering of IR photons off interstellar dust, for the effects of the Doppler shifting of decay lines from their rest wavelength associated with the Sun's velocity in the Galactic frame, and for the Doppler broadening of the line width arising from the nontrivial DM velocity dispersion, which are at a level resolvable by JWST in some data-taking modes. In Sec.~\ref{sec:Data_selection_and_processing}, we describe our data reduction pipeline, where we perform a custom processing and calibration using the official \texttt{jwst} code to avoid standard background subtractions which have the potential to remove a DM-induced spectral line from the data. In Sec.~\ref{sec:Likelihood_analysis_framework}, we describe our likelihood analysis framework in which we model the complicated IR backgrounds with non-parametric Gaussian Process (GP) modeling as pioneered for DM spectral line searches in astrophysical datasets in \cite{Foster:2021ngm, Foster:2022fxn}, and validate our framework through a direct re-analysis of the GN-z11 observations of Ref.~\cite{Janish:2023kvi}. Finally, in Sec.~\ref{sec:Results}, we present the results of our analysis, which we contextualize through comparison to the projections of \cite{Janish:2023kvi, Roy:2023omw} and previous analyses in \cite{Pinetti:2025owq, Saha:2025any}, through Fisher-information bounds on the sensitivities attainable from previously analyzed datasets, and through a dedicated analysis of archival PRISM observations, before making some concluding remarks in Sec.~\ref{sec:Conclusion}.

The main results of this work are supported by a number of appendices. In App.~\ref{app:SignalInjectionTests}, we present signal injection tests, in which synthetic signals are injected atop real data, that validate the performance of our analysis framework. In App.~\ref{app:SystematicAnalysisVariations}, we present results associated with systematic variations of our analysis framework that demonstrate the robustness of our claimed sensitivities with respect to analysis choices. In App.~\ref{app:KnownLines}, we describe the construction of our known-line candidate list and tabulate the astrophysical lines positively identified by our analysis. In App.~\ref{app:line_model_systematics}, we present alternate analyses which systematically vary our treatment of dust extinction in the line model, finding that our constraints vary at only the $\mathcal{O}(10\%)$ level. In App.~\ref{app:BackgroundSubtraction}, we demonstrate that our derived sensitivities are likewise insensitive to the treatment of the JWST-default background-subtraction steps. Finally, in App.~\ref{app:ModelIndependent}, we recast our  constraints as model-independent limits on the DM decay rate to photons.

In total, we argue that the decaying DM sensitivities which we develop, summarized in Fig.~\ref{fig:money_plot}, are highly robust. We make our  data reduction and analysis pipeline publicly available \href{https://github.com/joshwfoster/DarkJWST}{\faGithub}, enabling careful examination and reproduction of this work, as well as future re-analyses should improvements in instrumental modeling and data reduction make improved sensitivity attainable.

\section{IR lines from dark matter decays}
\label{sec:IR_Lines_Review}

In this Section, we outline our model for the DM decay signal. The high spectral resolution enabled by JWST, of order $10^{-3}$ or better, requires us to carefully model the line profile, which inherits its morphology from the DM velocity distribution in the barycentric rest frame (note that the data observed on-board JWST are converted to the barycentric rest frame during calibration). Furthermore, the IR photons produced from the DM decay suffer extinction due to scattering and absorption on the dust of the interstellar medium. Our analysis is the first search for DM decay in JWST data to simultaneously model both phenomena.

\subsection{Modeling DM decay lines with extinction}
\label{sec:extinction}

Axion DM of mass $\m$ decays to two photons at the rate $\Gamma = \g^2\m^3/64\pi$, generating a line flux proportional to the DM column density along the line of sight. This is characterized by the usual D-factor
\begin{equation}
    D = \int ds\, \rho(s),
    \label{eq:DFactor}
\end{equation}
where $\rho(s)$ is the axion DM density at distance $s$ along the line of sight at Galactic coordinates $(\ell, b)$, with the integration performed along all positive $s$. In the absence of extinction, the total line intensity is $\Phi = \Gamma D / 4\pi$. The IR photons produced in DM decays are additionally scattered and absorbed by interstellar dust, which attenuates the flux reaching JWST. Denoting by $A_\lambda(s)$ the integrated dust extinction for a photon of wavelength $\lambda$ emitted at distance $s$, we define the wavelength-dependent effective $D$-factor
\begin{equation}
    D_\mathrm{eff}(\lambda) = \int ds\, \rho(s)\, 10^{-0.4 A_\lambda(s)}.
    \label{eq:effective_DFactor}
\end{equation}
For the spatially homogeneous DM velocity distributions considered in this work, the flux spectrum expected at Galactic coordinates $(\ell,b)$ is then
\begin{equation}
    \frac{d\Phi}{d\nu}(\lambda;\ell,b) = \frac{\Gamma}{\m}\, D_\mathrm{eff}(\lambda) \left|\frac{df}{dv_{||}}\right|,
    \label{eq:decay_flux}
\end{equation}
where $df/dv_{||}$ is the DM line-of-sight velocity distribution evaluated at $v_{||} = 1 - 4\pi/ \lambda \m$.  For derivations, see {\it e.g.} Refs.~\cite{Dessert:2023vyl,Roy:2023omw}.

In our fiducial analysis, we assume that the DM density $\rho$ follows an NFW profile~\cite{Navarro:1995iw,Navarro:1996gj} with scale radius $r_s=24\,\mathrm{kpc}$ and normalization $\rho_s = 0.18\,\mathrm{GeV}/\mathrm{cm}^3$, which corresponds to a local DM density of $0.30\,\mathrm{GeV}/\mathrm{cm}^3$ with the Sun at $r_\odot=8\,\mathrm{kpc}$, identical to \cite{Saha:2025any}. JWST observes roughly $10^{-10}$ sr at a time, so the column density does not vary over the detector field-of-view.
The dust extinction reddens the observed spectrum due to its wavelength dependence. For a recent review, see~\cite{Draine:2003if}. The wavelength-dependence is approximately separable from its spatial-dependence: $A_\lambda(s) \approx A_{0.75\,\mu{\rm m}}(s) (\lambda/0.75\,\mu{\rm m})^{-1.85}$ away from the Galactic plane as determined by recent measurements of the near-infrared extinction law~\cite{2024ApJ...963...59B}. The wavelength-dependence can differ for lines-of-sight along the Galactic plane and near the GC, but we find that off-plane blank-sky locations, rather than the plane and the GC, are those which drive our sensitivity. The extinction is typically reported in units of magnitudes of extinction in the visible range, $A_V$, which we convert to $A_{0.75\,\mu{\rm m}}$ using the $R_V = 3.1$ extinction law of Ref.~\cite{1999PASP..111...63F}, as implemented in the Python package \texttt{dust\_extinction}~\cite{Gordon2024}. To correctly estimate the signal size, especially along the Galactic plane, we require measurements of $A_V(s)$ out to distances of order the scale radius. At the moment, such data exist only out to a few kpc and on a subset of the sky~\cite{2006A&A...453..635M,2019ApJ...887...93G}, while full-sky observations~\cite{1998ApJ...500..525S} have precisely measured $A_V(\infty)$, the extinction felt by a photon originating outside of the Galaxy.

In our fiducial analysis, we use the phenomenological \texttt{Drimmel03} extinction model of Ref.~\cite{2003A&A...409..205D}, which models 3-dimensional dust structure in the Milky Way, concentrated mostly in the spiral arms, and is built to roughly reproduce $A_V(\infty)$. This is implemented through the Python package \texttt{mwdust}~\cite{2016ApJ...818..130B}. We illustrate the effect of extinction at two representative masses in Fig.~\ref{fig:data_selection}. In App.~\ref{app:line_model_systematics}, we show how our results change if we replace this model with the \texttt{Combined19} map, which uses the data-driven results of Refs.~\cite{2006A&A...453..635M,2019ApJ...887...93G} where they are available; if we instead apply a uniform correction of $A_V(\infty)$ throughout the Galaxy with the \texttt{SFD} map of Ref.~\cite{1998ApJ...500..525S}; and if we replace the power-law wavelength-dependence of the reddening with the measured $0.09$--$32\,\mu\mathrm{m}$ Milky Way average extinction curve of Ref.~\cite{2023ApJ...950...86G}, which includes the silicate absorption features that fall within the MIRI band.

\subsection{Modeling the instrumental response to lines}
This decay flux arrives at JWST, where it is passed through the optics, which have finite spectral resolution. To model the signal as seen in the data, we convolve the intrinsic signal with the spectral response function (SRF) of the appropriate detector, which we approximate as a Gaussian with wavelength dispersion $\sigma_{\lambda,\rm inst}$. This approximation follows previous analyses and projections for DM decay in JWST data~\cite{Janish:2023kvi, Roy:2023omw}. For our fiducial velocity distribution, a homogeneous and isotropic Maxwell--Boltzmann distribution with characteristic velocity $v_0 = 220$ km/s~\cite{Evans:2018bqy}, the intrinsic Doppler-broadened line is also Gaussian, with wavelength dispersion
\begin{equation}
    \sigma_{\lambda,\rm DM}
    =
    \lambda_a \frac{\sigma_v}{c},
\end{equation}
where $\sigma_v = v_0/\sqrt{2} \simeq 156$ km/s is the line-of-sight velocity dispersion and $\lambda_a$ is the observed central wavelength of the line. The line center is additionally Doppler shifted by the solar motion, to $E = (\m/2)\,(1+\hat{n}\cdot\vec{\beta}_\odot)$, where $\hat{n}$ is the line-of-sight unit vector and $\vec{\beta}_\odot$ is the solar velocity in Galactic coordinates, fixed to its central value~\cite{2000A&A...354..522M, 2010MNRAS.403.1829S}. The total signal width is therefore
\begin{equation}
    \sigma_{\lambda,\rm sig}^2
    =
    \left(\lambda_a\frac{\sigma_v}{c}\right)^2
    +
    \sigma_{\lambda,\rm inst}^2.
    \label{eq:line_width}
\end{equation}
Throughout this work, we primarily characterize the linewidth by its full width at half maximum (FWHM), denoted $\Delta\lambda_{\rm sig}$. Under the Gaussian line-profile approximation,
\begin{equation}
    \Delta\lambda_{\rm sig}
    =
    2\sqrt{2\ln 2}\,\sigma_{\lambda,\rm sig}
    \simeq
    2.355\,\sigma_{\lambda,\rm sig}.
    \label{eq:FWHM}
\end{equation}
We use this total FWHM to define the characteristic width of the signal throughout the analysis.

For NIRSpec, we model the resolving power $R(\lambda)$ of each disperser, including the prism, by interpolating the dispersion curves distributed with the official JWST calibration reference data. For MIRI, we adopt the linear model $R(\lambda) = 4603 - 128\,(\lambda/\mu\mathrm{m})$~\cite{2023A&A...675A.111A,2023MNRAS.523.2519J}. In both cases, we take $\sigma_{\lambda,\rm inst} = \lambda_a/(2\sqrt{2\ln 2}\,R(\lambda_a))$, evaluating the resolving power at the central wavelength of the line, and we apply the instrumental response as a Gaussian convolution of the intrinsic spectrum computed on the wavelength grid of each dataset.

\section{Data selection and processing}
\label{sec:Data_selection_and_processing}

Our analyses make use of the public data \cite{mast_portal} collected from two JWST spectrographs: the Near-Infrared Spectrograph (NIRSpec)~\cite{2022A&A...661A..80J} and the Mid-Infrared Instrument (MIRI)~\cite{2015PASP..127..584R, 2015PASP..127..646W}. While archival data is available for a number of different observing modes, we focus on data collected in integral field unit (IFU) spectroscopy mode. The observations provide highly resolved flux density spectra as a function  of wavelength that are well-suited to our search for a spectral line feature. While other modes can reach comparable spectral resolution, the IFU observations provide calibrated surface-brightness spectra of the full observed field, which are directly suited to a signal that is spatially homogeneous over the field of view.

A particular danger for searches for decaying DM in the Milky Way halo is that although the decay may be spectrally identifiable as a line-like feature which appears above an otherwise smooth continuum background, the surface brightness associated with the decay will appear spatially homogeneous over the field of view of JWST. Data-driven background corrections which subtract away spatially homogeneous contributions to the observed data may then accidentally subtract away a genuine signal appearing in the data. We therefore seek to avoid application of data processing steps which could improperly remove evidence for a potentially DM-induced signal. 

\begin{figure*}[!htb]
    \centering
    \includegraphics[width=\linewidth]{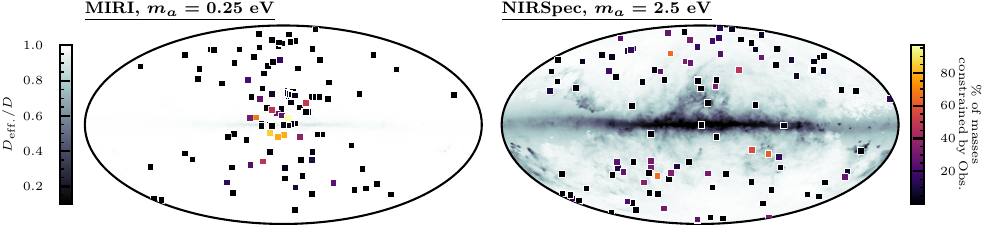}
    \caption{A joint illustration of the effective $D$-factor for two representative masses and the on-sky location of highly constraining datasets for MIRI and NIRSpec. \textbf{Left panel:} The sky-map shows the ratio $D_\mathrm{eff} / D$ for a mass of $m_a = 0.25 \,\mathrm{eV}$, a mass within the MIRI sensitivity range. Dark regions correspond to on-sky locations where extinction is an important effect. We find that under our fiducial model, extinction is a relatively unimportant effect, but see App.~\ref{app:line_model_systematics} for variations. Markers indicate the on-sky location of MIRI observations which are highly constraining in the sense that, for at least one mass considered in the analysis, the observation is among the 25 most constraining. Markers are colored by the fraction of analyzed masses for which the observation ranks among the 25 most constraining: darker markers correspond to observations relevant for only a small fraction of the masses considered in this work, while lighter markers correspond to observations relevant for a large fraction. \textbf{Right panel:} As in the left panel, but for a mass of $2.5\,\mathrm{eV}$, a mass that falls within the NIRSpec sensitivity range, and with markers corresponding to the location of constraining NIRSpec datasets. In this case, because extinction is a more important effect, fewer observations along the Galactic plane, where extinction is most significant, are among the most constraining datasets.}
    \label{fig:data_selection}
\end{figure*}

In this section, we describe our data reduction procedure for NIRSpec data in Sec.~\ref{sec:NIRSpec_Data_Reduction} and for MIRI data in Sec.~\ref{sec:MIRI_Data_Reduction}, and we compare our archival sample to those used by previous analyses in Sec.~\ref{sec:archival_comparison}. These data form the basis of our fiducial analysis, while the impact of alternative background-subtraction choices on our limits and detection significances is examined in App.~\ref{app:BackgroundSubtraction}. Our custom reductions of the NIRSpec and MIRI data use version \texttt{1.20.2} of the \texttt{jwst} package. The NIRSpec and MIRI samples contain all public data available as of July 14, 2026, restricted to observation-level (o-type) Stage 3 data products and excluding prism observations from the fiducial analysis. In Fig.~\ref{fig:data_selection}, we illustrate the on-sky locations of the most constraining datasets considered in this work, which are in part determined by extinction, see Sec.~\ref{sec:extinction}.

\subsection{NIRSpec data reduction}
\label{sec:NIRSpec_Data_Reduction}

For NIRSpec observations which require custom processing, our fiducial spectra are generated by rerunning the JWST calibration pipeline beginning from the Stage 0 uncalibrated FITS products. In these cases, we process each exposure through the standard Stage 1, Stage 2, and Stage 3 spectroscopic pipelines, with the modifications described below:
\begin{itemize}
    \item In Stage 2, we run the \texttt{Spec2Pipeline}, but disable the \texttt{bkg\_subtract} step. This step would otherwise subtract an image-by-image background estimate constructed from dedicated background exposures or nodded positions. Since this subtraction is designed to remove spatially smooth emission, it can also remove a diffuse line signal from decaying DM in the Milky Way halo.
    \item In Stage 3, we run the \texttt{Spec3Pipeline}, but disable the \texttt{master\_background} step. This step constructs a two-dimensional background model from a one-dimensional reference spectrum and subtracts it from the exposure before the final spectrum is extracted. As with image-by-image subtraction, this procedure can remove spatially smooth line emission and is therefore omitted in our fiducial reduction.
\end{itemize}
The \texttt{bkg\_subtract} and \texttt{master\_background} steps are applied only on a case-by-case basis in the public reductions. Out of the 1,690 unique Stage-3 NIRSpec spectra publicly available at the time we accessed the data archive, 187 had \texttt{bkg\_subtract} applied during Stage 2 and 2 had \texttt{master\_background} applied during Stage 3, as summarized in Table~\ref{tab:nirspec_background_subtraction}. For the 1,501 spectra for which neither step was applied, no custom processing is necessary, and we download the final Stage 3 data products directly from the archive. Only the remaining 189 are regenerated by reprocessing from the Stage 0 products with the background-subtraction steps disabled. We demonstrate in App.~\ref{app:BackgroundSubtraction} that our derived sensitivities are insensitive to these processing choices.

\begin{table}[t]
\centering
\small
\setlength{\tabcolsep}{4pt}
\begin{ruledtabular}
\begin{tabular}{lrrrr}
Source type & Neither & Image & Master & Total \\
\texttt{EXTENDED} & 869  & 82  & 2 & 953  \\
\texttt{UNKNOWN}  & 417  & 32  & 0 & 449  \\
\texttt{POINT}    & 215  & 73  & 0 & 288  \\
Total             & 1501 & 187 & 2 & 1690 \\
\end{tabular}
\end{ruledtabular}
\caption{
Summary of sky-background subtraction steps applied in the publicly available
NIRSpec reductions for the Stage-3 spectra used in this work. ``Neither'' denotes
observations for which neither \texttt{bkg\_subtract} nor
\texttt{master\_background} was applied. ``Image'' denotes image-by-image
background subtraction through \texttt{bkg\_subtract}, and ``Master'' denotes
master-background subtraction through \texttt{master\_background}.}
\label{tab:nirspec_background_subtraction}
\end{table}

From the resulting Stage 3 data products, whether directly downloaded or custom processed, we extract a one-dimensional background surface-brightness spectrum for each observation. For observations classified as \texttt{EXTENDED} or \texttt{UNKNOWN}, this background spectrum is naturally associated with the surface brightness over the observed field. For observations classified as \texttt{POINT}, the default source extraction is centered on the target aperture, and a background spectrum is only produced when a local-background region is defined. We therefore include \texttt{POINT} observations by extracting the local background from an annulus around the source. In the fiducial reduction, we use an annulus with inner and outer radii of $0.5''$ and $1.5''$, respectively, which masks the central source while retaining a large fraction of the surrounding field. For 32 of the 1,690 spectra, all of which were classified as \texttt{UNKNOWN}, neither the publicly available reduced data nor our custom reduction generated a nonzero background surface-brightness spectrum; these are excluded, leaving 1,658 NIRSpec spectra in the analysis sample.

\subsection{MIRI data reduction}
\label{sec:MIRI_Data_Reduction}

In contrast to our treatment of the NIRSpec data, our MIRI sample does not typically require a custom reprocessing. MIRI IFU observations are collected with the Medium Resolution Spectrometer (MRS), which covers wavelengths between approximately $4.9$ and $27.9\,\mu\mathrm{m}$. In the standard processing which generates the archival data products, neither the image-by-image \texttt{bkg\_subtract} step nor the \texttt{master\_background} step is normally applied to MRS science data, even for observations with associated dedicated background exposures~\cite{jdox_mrs_known_issues}. As a result, the archival Stage 3 data products generally retain the spatially homogeneous emission in which a DM decay signal would appear, and we may use them directly.

We download the final Stage 3 MRS data products \cite{mast_portal} and construct our spectra from the pipeline-produced one-dimensional extractions. For observations classified as \texttt{EXTENDED} or \texttt{UNKNOWN}, we use the extracted surface-brightness spectrum of the observed field. For observations classified as \texttt{POINT}, the default MRS extraction measures a local background in an annulus surrounding the source, and we take this pipeline-produced local-background spectrum as our data. In total, our primary MIRI sample contains 1,704 unique observation--target combinations, corresponding to 18,187 Stage-3 MRS spectra across the 12 MIRI channel and sub-band combinations.  A total of 20 atypical data products did have \texttt{master\_background} subtraction applied, requiring their custom processing, along the lines of the procedure in Sec.~\ref{sec:NIRSpec_Data_Reduction}.

\subsection{Comparison of archival datasets}
\label{sec:archival_comparison}

The public JWST archive grows continually, so analyses performed at different times draw on different volumes of data. Our sample contains all public NIRSpec and MIRI IFU observations available as of July 14, 2026, amounting to 1,214 NIRSpec and 1,704 MIRI unique observation--target combinations, corresponding to 1,690 NIRSpec and 18,187 MIRI Stage-3 spectra. By comparison, the analyses of Refs.~\cite{Pinetti:2025owq, Saha:2025any} accessed the archive roughly one year earlier, while the projections of Refs.~\cite{Janish:2023kvi, Roy:2023omw} predate the accumulation of most of the present archive. Differences in archival content cannot, however, account for the discrepancies among previously reported sensitivities examined in Sec.~\ref{sec:Results}: the coupling sensitivity scales only as the fourth root of the total exposure in the statistics-limited regime, and as we demonstrate in Sec.~\ref{sec:data_subselection}, the sensitivities attainable in practice are limited by systematic mismodeling rather than by the volume of available data.

Beyond the epoch of archival access, our data selection differs from those of previous analyses in two notable ways. First, Ref.~\cite{Saha:2025any} restricted its NIRSpec sample to observations classified as \texttt{EXTENDED} or \texttt{UNKNOWN}, for which the standard reduction directly provides a background spectrum of the observed field. By extracting local-background spectra for \texttt{POINT} observations within the same reduction framework, as described in Sec.~\ref{sec:NIRSpec_Data_Reduction}, we are able to include these observations in the fiducial analysis as well. In our sample, this adds 288 NIRSpec spectra. Second, we omit prism observations from our fiducial analysis because of their comparatively poor spectral resolution. Since prism data were included in the analysis of Ref.~\cite{Pinetti:2025owq}, we nevertheless examine their nominal information content in Sec.~\ref{sec:statistics_floor} and their analyzability in Sec.~\ref{sec:prism_analysis}. Conversely, a direct cross-match of the two samples shows that the only observations entering the analysis of Ref.~\cite{Pinetti:2025owq} which are absent from our sample are prism observations, which we exclude by construction, and candidate-type (c-type) association products, which are alternative pipeline combinations of archival exposures rather than additional observations.

Finally, care is required when comparing the number of datasets analyzed across works. The data release accompanying Ref.~\cite{Pinetti:2025owq} contains 3,639 NIRSpec and 15,419 MIRI flux density spectra. These totals, however, mix Stage 3 data products, which combine the exposures of an observation at the association level, with Stage 2 products extracted from the individual exposures themselves. Only 362 of the NIRSpec spectra and 2,974 of the MIRI spectra are Stage 3 products, while approximately $96\%$ and $99\%$, respectively, of the remaining Stage 2 spectra are extracted from the very exposures which enter the Stage 3 associations included in the same release.

The small remainder are Stage 2 spectra from observations for which no Stage 3 product appears in the release, including dedicated background pointings, absolute flux-calibration standards, and solar-system moving targets. These orphaned exposures amount to only 16 hours of NIRSpec and 10 hours of MIRI integration time, and so cannot meaningfully affect this accounting. Lacking Stage 3 products, they are excluded from our selection by construction; the moving targets among them are in any case poorly suited to a search for emission fixed in Galactic coordinates, as the line of sight sweeps across the sky over the course of the observation.

Counts which do not distinguish between processing stages therefore overstate the number of unique datasets by factors of roughly ten for NIRSpec and five for MIRI, and the tabulated exposure totals of 1,064 and 3,770 hours overstate the unique integration time by a factor of approximately two, as each exposure enters once through its Stage 2 products and again through its Stage 3 association. An analysis treating all of these spectra as independent thus assigns each exposure twice its statistical weight, artificially strengthening the inferred coupling sensitivity by a modest factor of $2^{1/4} \approx 1.2$, far too small to account for the order-of-magnitude difference in coupling sensitivity between Ref.~\cite{Pinetti:2025owq} and both Ref.~\cite{Saha:2025any} and this work.

Our restriction to o-type Stage-3 products deserves brief comment, since the archive also provides c-type Stage-3 products. A c-type association is a proposal-defined grouping, such as a mosaic or a set of repeated observations of a common target, from which the pipeline generates additional Stage-3 products by recombining exposures which already belong to the o-type associations. Such products therefore contain no additional data. Moreover, because a single observation may enter multiple candidate associations, there is no selection which admits c-type products without multiply counting exposures, short of auditing the constituent exposure lists of every product individually. The o-type products, by contrast, partition the archive uniquely: at fixed instrument configuration, each exposure contributes to exactly one product at any given wavelength. We therefore use o-type products exclusively so that no exposure is counted more than once at any given wavelength. The released dataset of Ref.~\cite{Pinetti:2025owq} illustrates the hazard of the alternative: its c-type products duplicate observations already present at the observation level, including one pair of spectra which are identical to one another (see Sec.~\ref{sec:floor_nirspec}).

\section{Analysis methodology}
\label{sec:Likelihood_analysis_framework}

\begin{figure}[!t]
    \centering
    \includegraphics[width=\linewidth]{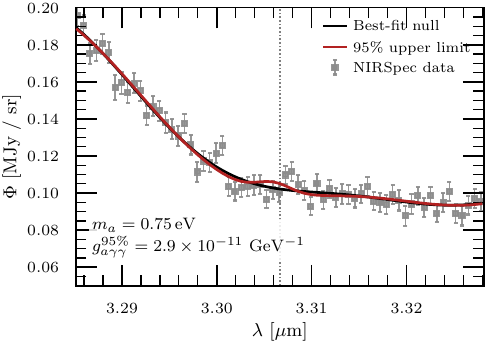}
    \caption{
    Illustration of the NIRSpec line analysis at $m_a = 0.75\,\mathrm{eV}$ using the individual dataset with the strongest expected constraining power at this mass, JWST observation \texttt{jw01964-o002\_t002} taken with the G395H grating. The points show the measured background spectrum with the rescaled uncertainties used in our fiducial analysis. The black curve shows the best-fit null model, while the red curve shows the profiled model evaluated at the 95\% CL upper limit, $g_{a\gamma\gamma}^{95\%}=2.9\times10^{-11}\,\mathrm{GeV}^{-1}$. The vertical dotted line indicates the expected axion-decay line wavelength at rest.
    }
    \label{fig:analysis_example}
\end{figure}

In this Section, we detail our analysis framework, which makes use of candidate DM decay lines modeled as described in Sec.~\ref{sec:IR_Lines_Review} and a combined parametric and nonparametric treatment of continuum backgrounds using a Gaussian process modeling treatment along the lines of those applied in \cite{Foster:2021ngm, Foster:2022fxn} as developed in \cite{Frate:2017mai}. We describe this procedure in Sec.~\ref{sec:gp_likelihood}, which we apply on a mass-by-mass basis for a set of masses defined in Sec.~\ref{sec:mass_by_mass}. In Fig.~\ref{fig:analysis_example}, we illustrate this analysis at a representative mass to which NIRSpec data are sensitive. We then describe in detail our procedure for quantifying and correcting systematic mismodeling effects via a data-calibrated spurious signal analysis in Sec.~\ref{sec:spurious_signal}, closely following previous applications in \cite{Foster:2021ngm, Salemi:2021gck}.

\subsection{Likelihoods with Gaussian process modeling}
\label{sec:gp_likelihood}

We begin by considering a dataset generated by the full processing pipeline as applied to a single JWST observation, which generates a calibrated observed flux density spectrum $\bm{d}$ and estimated statistical errors on the observed flux density $\bm{\sigma}$ at binned wavelengths $\bm{\lambda}$. We denote the values of the flux density spectrum, statistical error, and central value of the binned wavelengths in the $i^\mathrm{th}$ wavelength bin by $\bm{d}_i$, $\bm{\sigma}_i$, and $\bm{\lambda}_i$, respectively. Across datasets, the wavelength binning can vary significantly, associated with the resolving mode in which the data were collected and the details of the wavelength calibration, which can differ between observations even when performed within the same observing mode. We therefore choose to follow a joint likelihood analysis approach, in which each dataset is analyzed independently and subsequently combined at the level of the likelihood ratio, rather than a stacked analysis procedure.

For an observation which points at the on-sky location $(\ell, \,b)$, the decay flux can be calculated for a given axion mass $m_a$ following Eq.~\eqref{eq:decay_flux} of Sec.~\ref{sec:IR_Lines_Review}. We denote the canonically normalized decay flux density evaluated at the $i^\mathrm{th}$ wavelength bin for an axion with mass $m_a$ and a photon coupling $g_{a\gamma\gamma} = 10^{-10} \, \mathrm{GeV}^{-1}$ by $\bm{s}_i$. DM of mass $m_a$ decaying to two photons produces a line-like excess at $\lambda = 4 \pi/ m_a$ prior to Doppler and instrumental broadening. To perform a search in data, we therefore restrict our region of interest (ROI) only to data at wavelengths $\bm{\lambda}_i$ such that
\begin{equation*}
    \left| \bm\lambda_i - \frac{4 \pi}{m_a}\right|  \leq  5 \, \Delta \lambda_\mathrm{sig}
\end{equation*}
where $\Delta\lambda_\mathrm{sig}$ is the FWHM of the instrumentally broadened line defined in Eq.~\eqref{eq:FWHM}.

Within this ROI, we define the GP marginal likelihood for the data conditioned on the relevant model parameters by
\begin{widetext}
\begin{equation}
    \log \mathcal{L}(\bm{d} | A_\mathrm{sig}, \bm{\theta}, \bm{\theta}_\mathrm{GP}) = - \frac{1}{2} \left[ \bm{d} - \bm{\mu}(A_\mathrm{sig}, \bm{\theta})\right]^T\mathbf{K}^{-1} (\bm{\theta}_\mathrm{GP}) \left[ \bm{d} - \bm{\mu}(A_\mathrm{sig}, \bm{\theta})\right] - \frac{1}{2}\log|\mathbf{K}(\bm{\theta}_\mathrm{GP})|
\end{equation}
\end{widetext}
where $A_\mathrm{sig}$ is the normalization of the signal template serving as a proxy for $g_{a \gamma \gamma}^2$, $\bm{\theta}$ is the set of parametric background model parameters, and $\bm{\theta}_\mathrm{GP}$ is the set of Gaussian process model hyperparameters. The total parametric model $\bm{\mu}$ is comprised of the signal decay flux and a constant background continuum model as
\begin{equation}
\bm{\mu}(A_\mathrm{sig}, \bm{\theta} = \{B\}) = A_\mathrm{sig} \bm{s} + B.
\end{equation}
While our inference is performed in $A_\mathrm{sig}$, it is straightforwardly translated to $g_{a \gamma \gamma}$ by the relation
\begin{equation}
    A_\mathrm{sig} = \left(\frac{g_{a\gamma\gamma}}{10^{-10} \, \mathrm{GeV}^{-1}}\right)^2.
\end{equation}
Following standard practice, $A_\mathrm{sig}$ is allowed to take on negative values in our analysis, with their interpretation in terms of $g_{a\gamma\gamma}$ later prescribed by power-constraining procedures~\cite{Cowan:2011an}. 

We take the Gaussian process model kernel $\mathbf{K}$ to be modeled by the sum of a diagonal covariance corresponding to the pipeline-estimated statistical errors and an exponential-squared kernel such that
\begin{equation}
\begin{split}
\mathbf{K}_{ij}(&\bm{\theta}_\mathrm{GP}) \equiv \mathbf{K}(\bm\lambda_i, \bm{\lambda}_j, \bm{\theta}_\mathrm{GP} = \{A_\mathrm{noise}, A_\mathrm{GP} \}) \\ &=  A_\mathrm{noise}^2\bm\sigma_i^2 \delta_{ij} + A_\mathrm{GP}\exp \left[- \frac{(\bm{\lambda}_i - \bm{\lambda}_j)^2}{2 \sigma_\mathrm{GP}^2}\right].
\end{split}
\end{equation}
The Gaussian process kernel allows for a non-parametric modeling of features in the data which are smooth on the scale of $\sigma_\mathrm{GP}$. It is therefore critical that $\sigma_\mathrm{GP}$ not be allowed to take values which are as small or smaller than $\Delta \lambda_\mathrm{sig}$. In principle, this parameter could be treated as one free to be inferred by maximum likelihood estimation or to be profiled over. For simplicity, however, we fix $\sigma_\mathrm{GP} = 3\, \Delta \lambda_\mathrm{sig}$. We require $A_\mathrm{noise} > 0$ and $A_\mathrm{GP} \geq 0$, where $A_\mathrm{GP}$ carries the dimensions of a squared surface brightness; the background normalization $B$ is unconstrained. Values of $A_\mathrm{noise}$ below unity are permitted, since the Gaussian process covariance can trade off against the uncorrelated statistical uncertainties in describing the data.

For each dataset, we profile over the nuisance background parameters and Gaussian process hyperparameters to obtain
\begin{equation}
    \mathcal{L}(\bm{d} | A_\mathrm{sig}) \equiv \mathrm{max}_{\bm{\theta}, \bm{\theta}_\mathrm{GP}} \mathcal{L}(\bm{d} | A_\mathrm{sig}, \bm{\theta}, \bm{\theta}_\mathrm{GP}).
\end{equation}
To generate the likelihood over some large number of observations, we then evaluate
\begin{equation}
    \mathcal{L}(\{\bm{d}\} | A_\mathrm{sig}) = \prod_j \mathcal{L}(\bm{d}^{(j)} | A_\mathrm{sig})
    \label{eq:joint_likelihood}
\end{equation}
where $\bm{d}^{(j)}$ is the data associated with the $j^\mathrm{th}$ observation, accompanied by its own set of pipeline-estimated errors $\bm{\sigma}^{(j)}$ and calibrated wavelength bins $\bm{\lambda}^{(j)}$. We denote the value of $A_\mathrm{sig}$ which maximizes the profile likelihood by 
\begin{equation}
    \hat{A}_\mathrm{sig} \equiv \mathrm{argmax}_{A_\mathrm{sig}}\mathcal{L}(\{\bm{d}\} | A_\mathrm{sig})
\end{equation}
where the maximization is performed allowing for both positive and negative values of $A_\mathrm{sig}$.

With the joint-likelihood now defined in Eq.~\eqref{eq:joint_likelihood}, we define the standard test statistic for discovery by
\begin{equation}
    t = 2 \log \left[ \frac{ \mathcal{L}(\{\bm{d}\} | \hat{A}_\mathrm{sig}) }{\mathcal{L}(\{\bm{d}\} | A_\mathrm{sig} = 0) } \right].
    \label{eq:ts_for_discovery}
\end{equation}
When searching for evidence of a positive signal, the discovery test statistic $t$ is taken to be zero if the joint likelihood is maximized by a negative-valued $\hat{A}_\mathrm{sig}$. However, in our procedure for identifying and correcting for systematic mismodeling defined later in this section, we do not zero out the test statistic for discovery associated with negative best-fit $\hat{A}_\mathrm{sig}$ since positive and negative excesses are equally indicative of mismodeling. This test statistic can be directly related to the sign-weighted significance by
\begin{equation}
    Z = \mathrm{sign}(\hat{A}_\mathrm{sig}) \times \sqrt{t},
    \label{eq:sign_weighted_significance}
\end{equation}
which is expected to be normally distributed under the null. 

We also define the test statistic for upper limits
\begin{equation}
    q(A) = 2 \log \left[
    \frac{
        \mathcal{L}(\{\bm{d}\} | \hat A_\mathrm{sig})
    }{
        \mathcal{L}(\{\bm{d}\} | A_\mathrm{sig} = A)
    }
    \right].
    \label{eq:ts_for_limits}
\end{equation}
Assuming the asymptotic Wilks' theorem limit then allows us to determine a
one-sided 95\% confidence level (CL) upper limit on $A_\mathrm{sig}$ by
solving
\begin{equation}
    q(A_{\mathrm{sig},\,\mathrm{raw}}^{95})
    =
    \left[\Phi^{-1}(0.95)\right]^2 \approx 2.71,
\end{equation}
where $\Phi^{-1}$ is the inverse cumulative distribution function of the
standard normal distribution. To avoid reporting artificially strong limits from downward fluctuations, we
power constrain this result using the local curvature of the profiled
likelihood. Assuming uncertainties at the Cramér-Rao limit, the variance in estimators of $\hat{A}_\mathrm{sig}$ and $A_\mathrm{sig}^{95}$ is given by
\begin{equation}
    \sigma_A^{-2}
    =
    -
    \left.
    \frac{\partial^2 \log \mathcal{L}(\{\bm{d}\}|A_\mathrm{sig})}
    {\partial A_\mathrm{sig}^2}
    \right|_{A_\mathrm{sig}=\hat A_\mathrm{sig}}.
    \label{eq:parameter_estimation}
\end{equation}
We then impose a floor corresponding to a one-sigma downward fluctuation of
the expected 95\% CL upper limit,
\begin{equation}
    A_\mathrm{sig}^{95}
    =
    \max\left[
        A_{\mathrm{sig},\,\mathrm{raw}}^{95},
        \left(\Phi^{-1}(0.95) - 1\right)\sigma_A
    \right].
    \label{eq:power_constrained_limit}
\end{equation}
following the standard power constrained limit procedure \cite{Cowan:2011an}. This also forces our reported limits on $A_\mathrm{sig} \propto g_{a \gamma\gamma}^2$ to be positive-valued, translating to real-valued constraints on the magnitude of $g_{a\gamma \gamma}$.

\subsection{Mass-by-mass analyses}
\label{sec:mass_by_mass}

In total, the procedure of Sec.~\ref{sec:gp_likelihood} specifies the analysis at a single mass point. To analyze the total collection of NIRSpec data, we repeat our analysis procedure at a set of masses $m_a^{(i)}$ defined by
\begin{equation}
    m_a^{(i+1)} = \left(1 + \frac{\Delta \lambda_\mathrm{sig} ^{(i)}}{ 4 \lambda^{(i)}}\right) m_a^{(i)}
\end{equation}
with the lowest mass set by the largest wavelength $\lambda_\mathrm{max} = 5.27 \,\mu\mathrm{m}$ accessed by the NIRSpec instrument as
\begin{equation}
m_a^{(i = 0)} = 4\pi / \lambda_\mathrm{max}
\end{equation}
and $\Delta \lambda_\mathrm{sig} ^{(i)}$ the width of the signal line for an axion of mass $m_a^{(i)}$, centered at wavelength $\lambda^{(i)} = 4\pi/m_a^{(i)}$, as seen in the highest resolution observing mode.  The largest mass considered is the first mass for which 
\begin{equation}
    m_a^{(i)} > 4 \pi / \lambda_\mathrm{min}
\end{equation}
where $\lambda_\mathrm{min} = 0.9\,\mu\mathrm{m}$ is the minimum wavelength accessed by NIRSpec. The choice of mass step size is motivated by safely overresolving the wavelength resolution of the instrument. In total, we consider 5,406 masses; we follow an identical procedure to determine the set of masses used for MIRI analyses, resulting in 5,447 total masses.

\subsection{Comparison with the previous GN-z11 analysis}
\label{sec:gnz11_comparison}

\begin{figure*}[!htb]
    \centering
    \includegraphics[width=\linewidth]{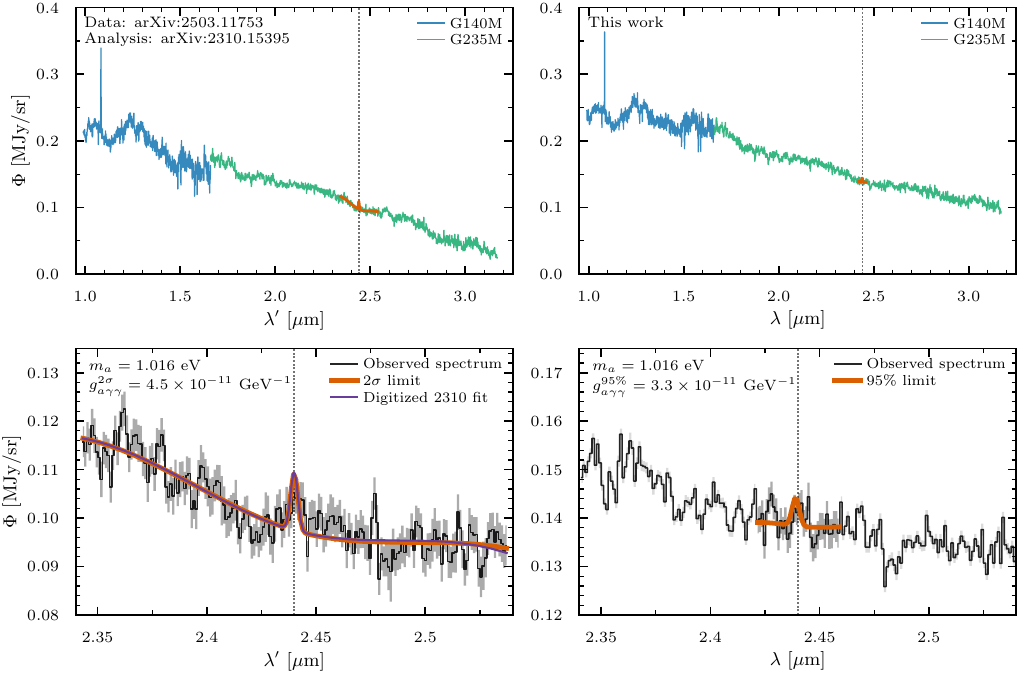}
\caption{A direct comparison of the analyses and associated physics results from our reproduction of the methods of \cite{Janish:2023kvi} and those developed and applied in this work. \textbf{Upper left:} the sky spectrum collected with the JWST NIRSpec instrument using filter/grating combinations of F100LP/G140M (blue) and F170LP/G235M (green) in observations of the target GN-z11 extracted from the public dataset provided by \cite{Pinetti:2025owq}. The spectrum is shown as a function of Galactic frame wavelength, indicated by $\lambda'$ (though see text). In orange, we illustrate a fit, using the public code of \cite{JanishCode}, for an axion decay line with a rest wavelength of $\lambda = 2.44\,\mu\mathrm{m}$, indicated by the dotted grey line, just as performed in \cite{Janish:2023kvi}. \textbf{Lower left:} A closer view of continuum and axion line contributions in the region of interest relevant to the spectral fit. Black indicates the sky spectrum, with the shaded grey indicating the data-driven error bars estimated following the methodology of \cite{Janish:2023kvi}. In orange, we provide our best-fit to the data for an axion with mass $m_a = 1.016$ eV and photon coupling $g_{a\gamma \gamma} = 4.5 \times 10^{-11}\,\mathrm{GeV}^{-1}$, which is the level associated with the $2\sigma$ upper limit on the coupling strength. In purple, we show the digitized fit of \cite{Janish:2023kvi}, shifted downward by $1.06\times10^{-3}\,\mathrm{MJy/sr}$ to account for the baseline difference between the data vintage of their published figure and the released data products (see text). The digitized fit demonstrates excellent consistency with our own, even though \cite{Janish:2023kvi} reports that their illustrated fit is for $g_{a\gamma \gamma} = 3.5 \times 10^{-11}\,\mathrm{GeV}^{-1}$. We believe this to be either a typo, or a confusion of their $2\sigma$ upper limit with the best-fit value of $g_{a \gamma \gamma}$, which we find coincidentally to be $3.5 \times 10^{-11}\,\mathrm{GeV}^{-1}$. See text for more discussion and validation. \textbf{Upper right:} As in the upper left, but for the data used in this work. While no custom processing was necessary, the data still differ somewhat from that provided in \cite{Pinetti:2025owq} because the public JWST data is reprocessed with updated reduction pipelines on a rolling basis. We work with the spectrum as a function of wavelength in the barycentric frame in which JWST data are natively provided; we indicate this by $\lambda$ rather than $\lambda'$. The axion+continuum fit in orange covers a comparatively smaller region because our analysis uses a smaller region of interest than in \cite{Janish:2023kvi}. \textbf{Lower right:} As in the lower left, but with the data used by and analysis performed in this work. Darker shaded regions illustrate error bars for data in the fit region, which are rescaled in a data-driven way, while lighter grey regions illustrate the error bars for data outside the fit region, which we do not rescale. The dotted grey line indicates the rest wavelength of the decay line, though the modeled line is Doppler shifted to somewhat shorter wavelengths. We find a 95\% CL upper limit on the axion-photon coupling of $3.3\times 10^{-11}\,\mathrm{GeV}^{-1}$, which is qualitatively consistent with the limit found in our reproduction of \cite{Janish:2023kvi}.}
    \label{fig:gnz11_comparison}
\end{figure*}

The first search for axion DM decay in JWST data was performed in Ref.~\cite{Janish:2023kvi}, which analyzed two NIRSpec IFU observations of the GN-z11 field, collected with the G140M/F100LP and G235M/F170LP disperser-filter combinations. The analysis code is public~\cite{JanishCode}, and in combination with the data release of Ref.~\cite{Pinetti:2025owq}, it provides a reproducible benchmark for validating our methodology. Unfortunately, the comparison cannot be made exact since archival JWST data products are periodically reprocessed as calibrations improve. By digitizing the figures of Ref.~\cite{Janish:2023kvi}, we find that the corresponding sky spectrum data products of Ref.~\cite{Pinetti:2025owq} differ slightly, though primarily by only a percent-level uniform baseline offset over the fit region of interest. By comparison, our more recently accessed data differ from those in Ref.~\cite{Janish:2023kvi} and Ref.~\cite{Pinetti:2025owq} by tens of percent. Working with the released products has the additional advantage that they are entirely independent of our own data downloading and processing, so that the comparison constructed here tests the analysis methodologies alone. However, we note that for these observations, no custom processing involving the deactivation of background-subtraction steps was necessary in any case, so the data product of our own pipeline is identical to that which would be downloaded directly from MAST, at least as of the time we accessed the archive.

In examining the analysis and modeling code of Ref.~\cite{JanishCode} and associated results in Fig.~1 of \cite{Janish:2023kvi}, we identify four errors or internal inconsistencies:
\begin{enumerate}
    \item The D-factor calculation evaluates trigonometric functions of
    Galactic coordinates expressed in degrees, yielding
    $D \approx 3.8\,\mathrm{GeV\,kpc\,cm^{-3}}$ for the GN-z11 sightline
    where the correct value is $D \approx 3.5\,\mathrm{GeV\,kpc\,cm^{-3}}$.
    \item The line-fitting window spans $150\,\sigma_v$, rather than $150$
    times the FWHM as described in the text.
    \item The Doppler correction from the barycentric to the
    Galactic frame is applied with an inverted sign. We denote by
    $\lambda'$ the resulting wavelength coordinate, shown in the left
    panels of Fig.~\ref{fig:gnz11_comparison}; at $2.44\,\mu\mathrm{m}$
    the modeled line centers are displaced by approximately $1.2$ times
    the total line width.
    \item The example line profile drawn in Fig.~1 of
    Ref.~\cite{Janish:2023kvi} is labeled as
    $g_{a\gamma\gamma} = 3.5\times10^{-11}\,\mathrm{GeV}^{-1}$ but
    corresponds to $g_{a\gamma\gamma} = 4.5\times10^{-11}\,
    \mathrm{GeV}^{-1}$, the $2\sigma$ upper limit obtained by their
    procedure at that mass. The best-fit coupling at this mass is
    coincidentally $g_{a\gamma\gamma} \approx 3.5\times10^{-11}\,
    \mathrm{GeV}^{-1}$, suggesting either a typographical error or a
    confusion between the best-fit and limiting values.
\end{enumerate}
For items 1--3, our reproduction follows the methodology implemented in the released code~\cite{JanishCode} rather than that described in the text~\cite{Janish:2023kvi}.

At masses for which Ref.~\cite{Janish:2023kvi} tabulates results, running their released pipeline on the spectra of \cite{Pinetti:2025owq} reproduces the tabulated limits provided in \cite{JanishCode} at the percent level. Item 4 cannot be validated in the same way, as the mass highlighted in their Fig.~1, $m_a = 1.016\,\mathrm{eV}$, is absent from the tabulated results. Instead, we refit the axion and continuum model with their methodology at this mass and compare directly with the published figure. For this comparison, the baseline offset identified above amounts to $1.06\times10^{-3}\,\mathrm{MJy\,sr^{-1}}$ over the fit window, and we shift the digitized curve downward by this amount in Fig.~\ref{fig:gnz11_comparison}. With this shift applied, the profile refit at $g_{a\gamma\gamma} = 4.5\times10^{-11}\,\mathrm{GeV}^{-1}$ reproduces the digitized published curve almost perfectly, as shown in the lower left panel.

The right panels of Fig.~\ref{fig:gnz11_comparison} show the analysis framework of this work applied to the same observations, now using the data products of our own pipeline, which as noted above reflect the current archival calibration. At $m_a = 1.016\,\mathrm{eV}$ we obtain a 95\% CL upper limit of $g_{a\gamma\gamma} = 3.3\times10^{-11}\, \mathrm{GeV}^{-1}$, qualitatively consistent with our reproduction of the limit obtained with the methodology of Ref.~\cite{Janish:2023kvi}. In comparing the two, we note that Ref.~\cite{Janish:2023kvi} reports $2\sigma$ limits rather than the 95\% CL limits used in this work; the distinction amounts to a percent-level difference in coupling and is not meaningful for any comparison performed here. Beyond this, the analyses differ in the signal model as well as in the statistical treatment. Our fit employs the line model of Sec.~\ref{sec:IR_Lines_Review}, which corrects the D-factor evaluation and the Doppler convention identified above and additionally accounts for the attenuation of the decay flux by dust extinction, though the extinction correction is small along this high-latitude sightline. Our fit is also performed in a considerably smaller region of interest, with the continuum modeled by a Gaussian process rather than a spline, and with the spectral errors rescaled by the data-driven procedure of Sec.~\ref{sec:gp_likelihood} rather than by a clipped residual standard deviation.

In total, this comparison validates our forward model and likelihood framework end-to-end on the one prior analysis which can be fully reproduced, and it establishes that the corrections identified above are minor for this dataset. In particular, none of the effects discussed here approach the order-of-magnitude discrepancies with Ref.~\cite{Pinetti:2025owq} we find in Sec.~\ref{sec:Results}.

\subsection{Systematic mismodeling corrections}
\label{sec:spurious_signal}

After a complete analysis of the data at all masses of interest, at each mass $m_a^{(i)}$, we have a test statistic in favor of discovery $t^{(i)}$. Under the null model, if the joint parametric and nonparametric Gaussian process treatment of the background were to perfectly describe the observed data at the level of the statistical uncertainties, Wilks' theorem implies that in the asymptotic limit,  the $t^{(i)}$ would follow a $\chi^2$ distribution with one degree of freedom. In practice, however, we find our prescribed background modeling procedure does not perform at this idealized level, making it necessary to implement a quantitative procedure to identify and correct for systematic mismodeling which may generate spurious excesses that might otherwise be interpreted as evidence in favor of decaying DM.

Following \cite{Salemi:2021gck}, we augment our joint likelihood with a spurious signal parameter
\begin{equation}
    \log \mathcal{L}(\{\bm{d}\}| A_\mathrm{sig}, A_\mathrm{nuis}) = \log \mathcal{L}(\{\bm{d}\}| A_\mathrm{sig} + A_\mathrm{nuis}) - \frac{A_\mathrm{nuis}^2}{2 \sigma_\mathrm{nuis}^2}.
    \label{eq:nuisance_augmented_likelihood}
\end{equation}
Formally, this corresponds to taking a mean-zero Gaussian prior with variance $\sigma_{\mathrm{nuis}}^2$ on the size of a line-like excess under the null. We then define the profiled-likelihood after augmentation
\begin{equation}
    \tilde{\mathcal{L}}(\{\bm{d}\} | A_\mathrm{sig}) = \mathrm{max}_{A_\mathrm{nuis}}\mathcal{L}(\{\bm{d}\}| A_\mathrm{sig}, A_\mathrm{nuis})
\end{equation}
with an associated test statistic for discovery $\tilde t$ and test statistic for upper limits $\tilde q$, defined identically to Eq.~\eqref{eq:ts_for_discovery} and Eq.~\eqref{eq:ts_for_limits}, respectively, with $\mathcal{L} \rightarrow \tilde{\mathcal{L}}$. 

In order to make use of these augmented test statistics for upper limits and discovery, an appropriate value of $\sigma_\mathrm{nuis}$ must be chosen. We make such a choice on a mass-by-mass basis in the following way. First, at mass $m_a^{(i)}$, we restrict interest to only masses $m_a^{(j)}$ such that
\begin{equation*}
\Delta \lambda_\mathrm{sig} ^{(i)} \leq | \lambda^{(i)} - \lambda^{(j)}| \leq 50\, \Delta \lambda_\mathrm{sig} ^{(i)},
\end{equation*}
which corresponds to masses that produce a line-like excess within 50 FWHMs of the line but not within a single FWHM of the line which would be produced by the mass of interest. We then consider the distribution of the $\{t^{(j)}\}$ associated with that restricted set of $\{m_a^{(j)}\}$. 

Because this total set of masses is roughly 100 FWHMs in width, we would expect roughly 100 statistically independent mass points, with $10\%$ of them possessing a test statistic for discovery which exceeds the $90^\mathrm{th}$ percentile value of $2.71$ expected in the Wilks' limit under the null. If fewer than $10\%$ of the masses exceed the threshold of $2.71$, then we choose $\sigma_\mathrm{nuis}$ associated with $m_a^{(i)}$ to be zero. Otherwise, we tune the variance which sets the nuisance signal prior to its minimum value such that only $10\%$ of the $\tilde t^{(j)}$ associated with the restricted set of masses exceed $2.71$. For the mass of interest, we then report the $\tilde t^{(i)}$ and 95\% CL upper limit determined from $\tilde q^{(i)}$ using the tuned value of $\sigma_\mathrm{nuis}$. We repeat this procedure at each mass, resulting in a data-driven correction for systematic mismodeling effects. As we demonstrate in greater detail in Sec.~\ref{sec:Results}, this has a significant impact on the DM decay limits developed in this work and colors our outlook on future searches for decaying DM with ongoing JWST data collections.

Throughout this work, expected limits and their one- and two-sigma containment intervals are derived from the local curvature of the profiled likelihood, following Eq.~\eqref{eq:parameter_estimation}. For results incorporating the spurious-signal correction, the curvature is that of the nuisance-augmented likelihood evaluated at the tuned value of $\sigma_\mathrm{nuis}$, so that the corresponding expected limits are conditional on the data-calibrated nuisance scale rather than pre-data expectations under a fixed model.

\begin{figure*}[!htb]
    \centering
    \includegraphics[width=.99\linewidth]{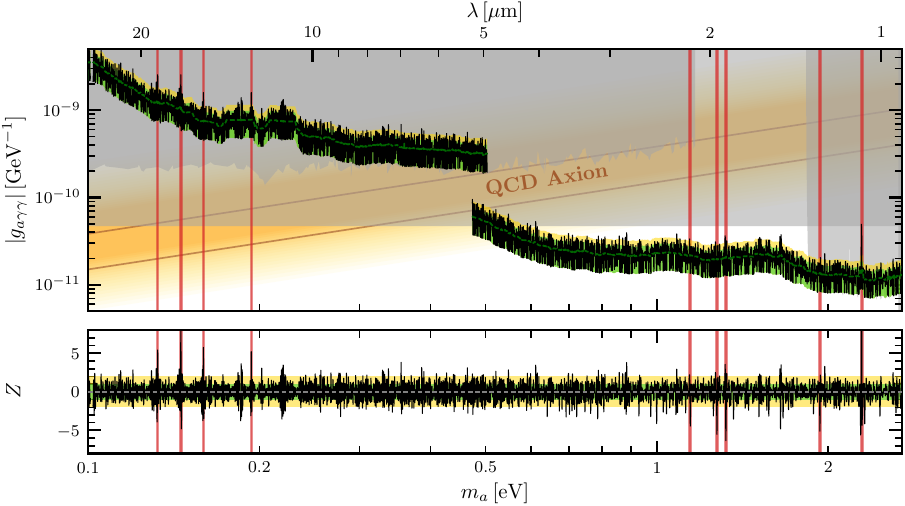}
    \caption{Results from our fiducial analysis of JWST data. \textbf{Top panel:} In black, we present the 95\% CL power-constrained upper limits on the axion-photon coupling $g_{a\gamma\gamma}$ as a function of mass. Expected one- and two-sigma containment intervals of the limit are depicted in green and gold, respectively, with the dashed green line corresponding to the median expected 95\% CL limit. Shaded red regions, with width equal to the instrumental resolution at that wavelength,  correspond to high-significance detections of known astrophysical IR lines associated with hydrogen, helium, neon, and sulfur. The derived limits realize a sharp transition at $m_a \approx 0.5\,\mathrm{eV}$, corresponding to the transition between masses which are probed by the comparatively lower spectral resolution MIRI instrument and the comparatively higher resolution NIRSpec instrument. Other constraints on this parameter space include the CAST helioscope~\cite{CAST:2007jps, CAST:2008ixs, CAST:2011rjr, CAST:2013bqn, CAST:2015qbl, CAST:2017uph} and globular cluster cooling \cite{Ayala:2014pea, Dolan:2022kul}, as well as searches for decay lines with WINERED \cite{Yin:2024lla} and MUSE \cite{Regis:2020fhw, Todarello:2023hdk}. \textbf{Bottom panel:} The sign-weighted detection significance, as defined in Eq.~\eqref{eq:sign_weighted_significance}, as a function of mass. As in the top panel, the identified astrophysical lines are illustrated with a red band. All were detected with $t \gg t_\mathrm{thresh}$ prior to the spurious-signal correction; the plotted post-correction significances are substantially reduced, though the strongest features remain above threshold and extend beyond the plotted range. No other high significance lines were identified. }
    \label{fig:money_plot}
\end{figure*}

\subsection{Look elsewhere effect}

Since our analysis considers many candidate masses at which a line-like signal might appear, correction for the look-elsewhere effect is relevant. In both NIRSpec and MIRI analyses, we consider approximately $5500$ masses, overresolving the linewidth by a factor of four. Summing over the NIRSpec and MIRI mass grids, we approximate our combined search as interrogating $2,750$ statistically independent possible lines. We therefore set the threshold for a $5\sigma$ detection by 
\begin{equation}
    t_\mathrm{thresh} = F_{\chi^2_1}^{-1} \big[ F_{\chi^2_1}(25)^{1/2750}\big] \approx 40
\end{equation}
where $F_{\chi^2_1}$ is the cumulative distribution function of the $\chi^2$-distribution with one degree of freedom and $F_{\chi^2_1}^{-1}$ its inverse, \textit{i.e.}, the percentile point function. This threshold is such that the probability of one of the $2,750$ total independent line locations realizing an excess (or deficit) with $t \geq t_\mathrm{thresh}$ is equal to the probability of a single possible line location realizing an excess or deficit with $t \geq 25$. As a result, $t_\mathrm{thresh}$ represents a $5\sigma$ detection threshold calibrated against the look-elsewhere effect for our analysis. This threshold assumes that the null distribution of the corrected test statistics follows the asymptotic $\chi^2_1$ form of Wilks' theorem into the deep tail. As we demonstrate in Sec.~\ref{sec:Results}, the ensemble of corrected test statistics is empirically consistent with this expectation over the range accessible to our mass grids, corresponding to survival fractions of a few $\times 10^{-4}$.

\section{Results}
\label{sec:Results}

The results of our fiducial analysis, described in detail in Sec.~\ref{sec:Likelihood_analysis_framework} and including the limit and test statistic corrections through the data-calibrated correction for systematic mismodeling via our spurious signal treatment, are presented in Fig.~\ref{fig:money_plot}. We find the NIRSpec data to result in the best constraints on $g_{a\gamma\gamma}$, at about the level of $g_{a\gamma\gamma}^{95} \approx \mathrm{few} \times 10^{-11} \,\mathrm{GeV}^{-1}$ across the full range of masses to which the instrument is sensitive. By comparison, the MIRI data realize generally worse sensitivity to decaying DM, resulting in constraints on $g_{a\gamma\gamma}$ ranging between $3 \times 10^{-10}\,\mathrm{GeV}^{-1}$ and $5 \times 10^{-9}\,\mathrm{GeV}^{-1}$. 

\begin{figure}[!htb]
    \centering
    \includegraphics[width=\linewidth]{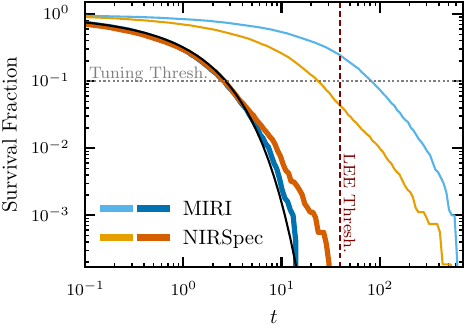}
    \caption{The survival functions of the discovery test statistic $t$ from the likelihood analysis of NIRSpec and MIRI data. Here, $t$ is evaluated without zeroing at negative $\hat{A}_\mathrm{sig}$, so that both excesses and deficits are retained. The $y$-axis indicates the fraction of mass points tested with a discovery test statistic at or above the value on the $x$-axis. Under the null, the $t$ would follow a $\chi^2$-distribution with one degree of freedom, corresponding to an expected survival function illustrated in black. The dashed red line indicates the threshold for a discovery test statistic corresponding to a LEE-calibrated detection at or above $5\sigma$ global significance. Prior to correction by the data-driven spurious signal treatment, the survival functions from MIRI (light blue) and NIRSpec (light orange) show no consistency with the expected survival function, with many high significance detections. The spurious-signal-corrected survival functions, for which known lines are masked out, are shown in dark blue and dark orange for MIRI and NIRSpec, respectively. After this correction, the agreement with the expected survival function is relatively good, and no high significance excesses persist.}
    \label{fig:survival_function}
\end{figure}

We also present the distribution of test statistics for discovery $t$ from the NIRSpec and MIRI analyses, both before and after spurious signal systematic correction in Fig.~\ref{fig:survival_function}. After the full joint analysis but before the application of the systematic mismodeling correction, we find that roughly $5\%$ of masses in the NIRSpec range and roughly $25\%$ of the masses in the MIRI range exceeded our LEE-calibrated detection threshold. Some of those high significance detections are associated with known astrophysical lines, see App.~\ref{app:KnownLines}. In applying our spurious signal correction, we first apply it blindly to the data, then identify any excess or deficit at or above $4\sigma$ local significance consistent with an expected line-like signal, then re-calibrate the spurious signal nuisance hyperparameter with these detections masked out. We also exclude them from our corrected survival function. After application of the mismodeling correction, we find no excesses above our LEE-calibrated detection threshold beyond the known astrophysical lines of App.~\ref{app:KnownLines}. The correction substantially reduces the significance of every identified line, though the strongest among them remain above threshold. We therefore report entirely null results for decaying DM from our analysis.

This behavior admits a simple interpretation. Prior to correction, a sizable fraction of the masses inspected realize line-like features at high local significance, whether from astrophysical lines leaking into an otherwise clean sample or from coherent instrumental structure. These features constitute a confusion limit: under the prior that at most one of them can be a DM decay line, the significance of any individual feature is meaningful only relative to the population of comparable features across the mass range. The spurious-signal correction reduces the significance of a genuine line by the same factor as the surrounding mismodeling, so that a DM line remains detectable provided it is as prominent as the astrophysical lines that survive the correction.

The sensitivities derived from NIRSpec following our fiducial analysis pipeline are roughly consistent with those found in \cite{Saha:2025any}, though approximately an order of magnitude weaker than those found in \cite{Pinetti:2025owq}. No MIRI analysis was presented in \cite{Saha:2025any}. Our discrepancy with the MIRI results of \cite{Pinetti:2025owq} is even greater than in the case of NIRSpec, corresponding to roughly two orders of magnitude. Since the axion decay rate scales as $g_{a \gamma \gamma}^2$, these differences imply a two order of magnitude and four order of magnitude inconsistency between the flux density sensitivities derived in this work and that of \cite{Pinetti:2025owq}.

We examine the agreement, or lack thereof, with prior works in great detail in the remainder of this section. We find generally excellent consistency between our analysis, the projections of \cite{Roy:2023omw, Janish:2023kvi}, and the analysis of \cite{Saha:2025any} after accounting for differences in exposure time and the inclusion of our data-driven noise estimates and spurious signal systematic correction, which have the effect of weakening our derived sensitivities but increasing the robustness of our analysis. We study these effects in detail in Sec.~\ref{sec:data_subselection}  and Sec.~\ref{sec:data_driven_errors}. On the other hand, we find the strong constraints of \cite{Pinetti:2025owq} to be non-reproducible, even when working with the publicly available dataset associated with that work. 

\subsection{Data subselection}
\label{sec:data_subselection}

\begin{figure*}[!htb]
    \centering
    \includegraphics[width=\linewidth]{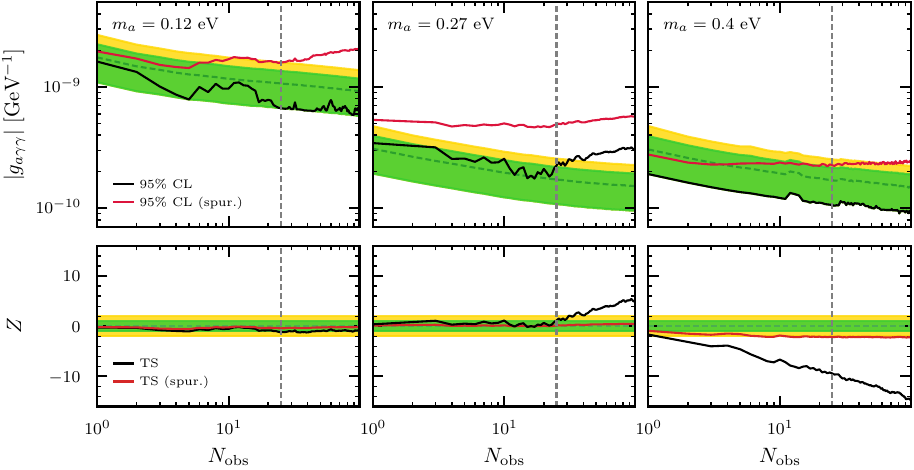}
    \caption{Dependence of the MIRI sensitivity on the number of observations included in the joint analysis, shown for three representative axion masses. Observations are included from strongest to weakest expected sensitivity. \textbf{Top panels:} The dashed green curves show the expected sensitivity obtained from the full joint analysis, with the green and gold bands denoting the corresponding one- and two-sigma containment intervals. The black curves show the observed limits prior to the spurious-signal correction, while the red curves show the final limits after the correction. \textbf{Bottom panels:} The corresponding sign-weighted discovery significances before (black) and after (red) application of the spurious-signal correction. A high significance excess (deficit) emerges for large $N_\mathrm{obs}$ at mass $m_a = 0.27\,\mathrm{eV}$ ($m_a = 0.4\,\mathrm{eV}$), while the $m_a = 0.12\,\mathrm{eV}$ analysis stays relatively well-behaved. After correction by the spurious signal treatment, none exhibit a significant excess, but the sensitivity worsens with increasing $N_\mathrm{obs}$ after $N_\mathrm{obs} \geq 25$, indicated by a dashed grey line, as increased data volume uncovers statistically significant mismodeling that must be corrected.}
    \label{fig:miri_data_subselection}
\end{figure*}

\begin{figure*}[!htb]
    \centering
    \includegraphics[width=\linewidth]{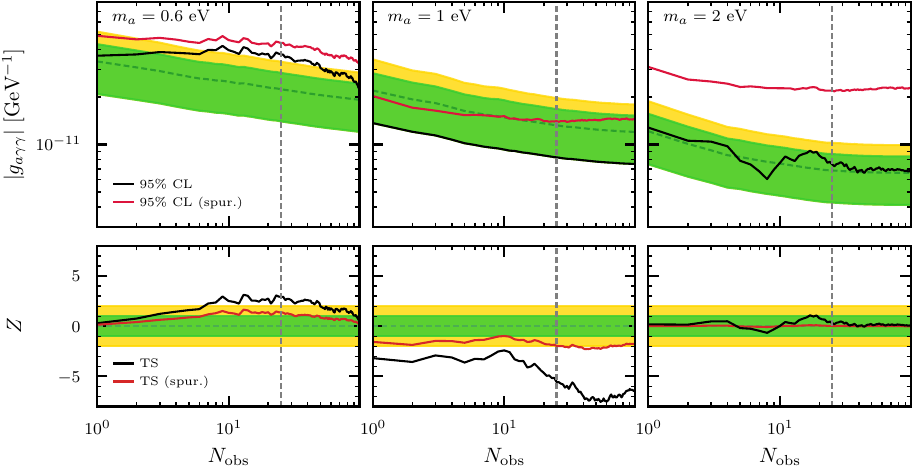}
    \caption{As in Fig.~\ref{fig:miri_data_subselection}, but for three representative masses within the NIRSpec sensitivity range. As in the case of MIRI, the systematic mismodeling correction results in none of these masses realizing a high significance excess, but the sensitivity tends to saturate at $N_\mathrm{obs} \gtrsim 25$.}
    \label{fig:nirspec_data_subselection}
\end{figure*}

The sensitivities derived in this work are fundamentally limited by the fact that increasing the total exposure through the joint analysis of an increasing number of datasets does not result in limits which improve with exposure as expected. To study this behavior, at each axion mass we independently analyze all publicly available observations and rank them from strongest to weakest according to their expected sensitivity, as determined from the local curvature of the profile likelihood following Eq.~\eqref{eq:parameter_estimation}. Because this expected sensitivity is evaluated about the best-fit model, allowing for a nonzero $\hat A_\mathrm{sig}$, the ranking is agnostic to the presence of a positive or negative line-like residual in an individual observation.

We then progressively construct joint analyses from the top $N_\mathrm{obs}$ observations in this ranking. The resulting behavior is illustrated for three representative masses in the MIRI and NIRSpec ranges in Figs.~\ref{fig:miri_data_subselection} and \ref{fig:nirspec_data_subselection}, respectively. The dashed green curves show the expected sensitivity obtained from the full joint analysis, in which the noise-rescaling and Gaussian-process hyperparameters are re-estimated as additional observations are incorporated. We have verified that this joint expected sensitivity agrees with the sensitivity expected from statistically combining the independently inferred constraining power of the included observations, each of which accounts for its own freely estimated noise-rescaling and Gaussian-process hyperparameters. The expected sensitivity therefore improves with the inclusion of additional data at the statistically anticipated rate.

The observed limits, however, do not realize this expected improvement. The black curves show that the joint analyses can develop substantial positive or negative line-like residuals as additional observations are included, with corresponding discovery significances well outside the range expected under the null. This residual mismodeling motivates the spurious-signal treatment of Sec.~\ref{sec:spurious_signal}. The corrected limits and significances are shown in red. While the correction substantially reduces the anomalous significances, it further weakens the resulting limits. Beyond a relatively modest number of the most sensitive observations, additional datasets therefore provide little improvement in the robust sensitivity to decaying DM. We therefore have constructed our fiducial joint likelihood from the 25 observations with the strongest expected sensitivity at each mass. However, in App.~\ref{app:SystematicAnalysisVariations}, we present alternative results in which we include only the top 10 and up to 50 datasets, demonstrating the robustness of our fiducial choice. 

This saturation carries an implication beyond the choice of our fiducial sample. Because the significance of the line-like mismodeling grows along with the statistical power of the joint analysis, the robust sensitivity of the archive is set not by its accumulated exposure but by the fidelity of the instrumental modeling and data reduction. The limits reported in this work are therefore unlikely to be improved upon by continued data taking or by re-analysis of the growing archive; meaningful gains will instead require reductions in the line-like mismodeling itself, at the level of the spectral uncertainties, through improved calibration and processing pipelines.

\subsection{Impact of data-driven error corrections}
\label{sec:data_driven_errors}

\begin{figure}[!htb]
    \centering
    \includegraphics[width=\linewidth]{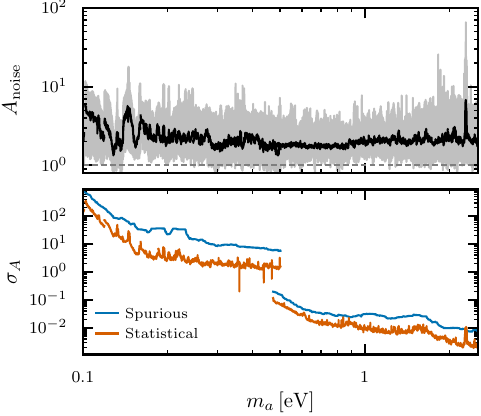}
    \caption{An illustration of the sensitivity-weakening inputs to our analysis as a function of mass. \textbf{Top panel:} The error-rescaling parameter $A_\mathrm{noise}$ for the 25 datasets entering our fiducial joint analysis at each mass. The solid black line is the median over the 25 datasets, and the shaded band spans their minimum and maximum. The median rescaling is approximately two across the NIRSpec range and rises toward five at the low-mass end probed by MIRI, while individual datasets can prefer rescalings of an order of magnitude or more. \textbf{Bottom panel:} The spurious-signal scale $\sigma_\mathrm{nuis}$ (blue, labeled Spurious) compared with the statistical uncertainty $\sigma_A$ on the signal normalization (orange, labeled Statistical). At nearly all masses, $\sigma_\mathrm{nuis}$ exceeds $\sigma_A$: the sensitivities are primarily determined by the mitigation of systematic mismodeling. See text for details.}
    \label{fig:error_analysis}
\end{figure}

Within our analysis, there are two effects which act to weaken the sensitivity of our joint analysis. The first is that although the JWST data reduction pipeline generates estimated statistical errors at each wavelength in the flux density spectrum, our analysis includes a nuisance parameter which multiplicatively rescales the error bars as they enter the GP likelihood, see Sec.~\ref{sec:gp_likelihood} for details. In the top panel of Fig.~\ref{fig:error_analysis}, we show the median and spread of the error-rescaling parameters $A_\mathrm{noise}$ for the 25 datasets used in our fiducial analysis as a function of mass. The data typically prefer error bars twice as large as those generated by the data reduction pipeline, and for individual datasets the preferred rescaling can exceed an order of magnitude. Because the sensitivity to $g_{a\gamma\gamma}$ scales with the square root of the flux density errors, the typical rescaling weakens the limit by a factor of approximately $\sqrt{2} \approx 1.4$.

The derived sensitivity is further weakened by the spurious signal nuisance parameter which corrects for systematic mismodeling. The necessity of an appreciable correction is made apparent by the fact that a large fraction of the masses at which we seek evidence for a line-like signal realize discovery test statistics exceeding the LEE-calibrated threshold prior to correction. To assess the size of the correction, the bottom panel of Fig.~\ref{fig:error_analysis} compares the anticipated statistical uncertainty $\sigma_A$ on the signal normalization $A_\mathrm{sig}$, evaluated from the data using Eq.~\eqref{eq:parameter_estimation} prior to nuisance augmentation (labeled Statistical), with the spurious-signal scale $\sigma_\mathrm{nuis}$ (labeled Spurious).

At nearly all masses, $\sigma_\mathrm{nuis}$ exceeds $\sigma_A$: systematic mismodeling is the dominant uncertainty of this analysis. In the NIRSpec range, $\sigma_\mathrm{nuis}$ is typically a factor of $2$--$3$ larger than $\sigma_A$, weakening the coupling sensitivity by a factor of approximately $1.5$; in the MIRI range, the ratio is typically $\sim 5$, weakening the coupling sensitivity by a factor of approximately $2$--$3$, consistent with the comparison of Fig.~\ref{fig:result_comparison}. Accounting for both effects, our sensitivities are roughly $2$--$4$ times weaker than would be expected from exposure time and the pipeline-reported statistical errors alone. 

\subsection{Comparison with other projections and limits}
\label{sec:result_comparison}

\begin{figure*}[!htb]
    \centering
    \includegraphics[width=\linewidth]{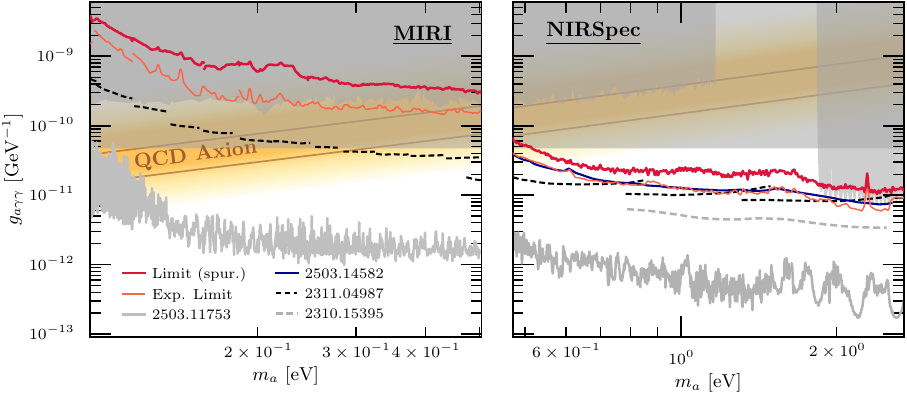}
    \caption{A comparison of our derived sensitivities with other projections and constraints derived in the literature. The gold band indicates the QCD axion parameter space, and shaded gray regions indicate existing constraints, following the conventions of Fig.~\ref{fig:money_plot}. \textbf{Left panel:} Here we consider the MIRI sensitivity to axion decay. The constraints developed under our fiducial analysis are shown in dark red and include the weakening effect of the spurious signal treatment; in light red, we illustrate the expected constraints prior to spurious signal treatment. In both cases, we have smoothed the results as a function of mass for clarity. This comparison reveals that the spurious signal correction for systematic mismodeling weakens the limits by roughly a factor of $3$ in the mass range probed by MIRI. We compare to the projected sensitivity of \cite{Roy:2023omw} for the 15-year lifespan of JWST, rescaled by a factor of $5^{1/4}$ to roughly account for the 3-year data volume considered in this work. Our limits are considerably weaker than these projections, primarily due to the data-driven error estimates and the failure of the data to accumulate sensitivity with increasing data volume as studied in Sec.~\ref{sec:data_subselection}. The constraints of \cite{Pinetti:2025owq} claim a sensitivity an order of magnitude beyond even this projection. \textbf{Right panel:} As in the left panel, but for the NIRSpec instrument. We additionally illustrate the derived sensitivities of \cite{Saha:2025any} (dark blue) and the projections of \cite{Janish:2023kvi} (dashed grey). Up to the weakening effect of our spurious signal treatment, which is smaller for NIRSpec than for MIRI, good consistency is observed between our results, the constraints of \cite{Saha:2025any}, and the projections of \cite{Roy:2023omw}, while \cite{Janish:2023kvi} is somewhat discrepant, projecting twice as strong constraints. As before, the results of \cite{Pinetti:2025owq} are an order of magnitude or more stronger than all other projections or constraints.}
    \label{fig:result_comparison}
\end{figure*}

The sensitivity of JWST blank-sky observations to decaying axion DM was first estimated in the projections of Refs.~\cite{Janish:2023kvi, Roy:2023omw}, which assumed statistics-limited analyses of archival exposure with pipeline-level uncertainties and well-modeled continua. In Fig.~\ref{fig:result_comparison}, we compare our fiducial limits with these projections.

Accounting for the difference between the archival exposure assumed in the projections of Ref.~\cite{Roy:2023omw} and that entering our joint analyses, we find our NIRSpec limits are weaker than a naive statistical expectation by an $\mathcal{O}(1)$ factor, fully consistent with the data-driven corrections quantified in Sec.~\ref{sec:data_driven_errors}. We observe a similar level of agreement with the analysis of Ref.~\cite{Saha:2025any}: prior to our correction for systematic mismodeling, our expected joint sensitivity agrees with their reported limits to within $10\%$, while our final corrected limits are weaker than theirs by $30$--$50\%$. Given the appreciable methodological differences, including our inclusion of \texttt{POINT}-classified observations, the custom data reduction, dust extinction modeling, and systematic mismodeling correction, we regard the observed level of agreement between fully independent pipelines as a meaningful cross-check of both analyses. By contrast, the projections of Ref.~\cite{Janish:2023kvi} anticipated somewhat greater sensitivity; the factor of two difference between the published NIRSpec projections of Ref.~\cite{Roy:2023omw} and Ref.~\cite{Janish:2023kvi} has remained unresolved, and is of lesser present-day importance given the availability of real JWST data. More strikingly, Ref.~\cite{Pinetti:2025owq} claims a further order of magnitude improvement in sensitivity compared to our results and all other analyses and projections. We examine that work directly through its GN-z11 predecessor in Sec.~\ref{sec:gnz11_comparison} and through the information content of its released dataset in Sec.~\ref{sec:statistics_floor}, finding that the strength of those claimed limits is incompatible with even the most optimistic sensitivity that could be anticipated from the datasets upon which they depended.

Fewer comparisons can be made in terms of MIRI since Ref.~\cite{Janish:2023kvi} made no projections, and Ref.~\cite{Saha:2025any} published no limits associated with the instrument. We find that our constraints are weaker by a factor of a few as compared to Ref.~\cite{Roy:2023omw}, though once again this is consistent with the various sensitivity-weakening effects relevant for this work. Additionally, through private correspondence with the authors, we have been informed that our MIRI results are approximately consistent with unpublished ones developed during production of Ref.~\cite{Saha:2025any}. Finally, as before, we find the claimed sensitivities of Ref.~\cite{Pinetti:2025owq} considerably stronger, amounting to nearly two orders of magnitude, and once again, when we inspect it more carefully in Sec.~\ref{sec:statistics_floor}, we find it to be incompatible with the statistics floor associated with the datasets considered in that work.

\subsection{Fisher-information bounds on attainable DM sensitivities}
\label{sec:statistics_floor}

\begin{figure*}[!htb]
    \centering
    \includegraphics[width=.99\linewidth]{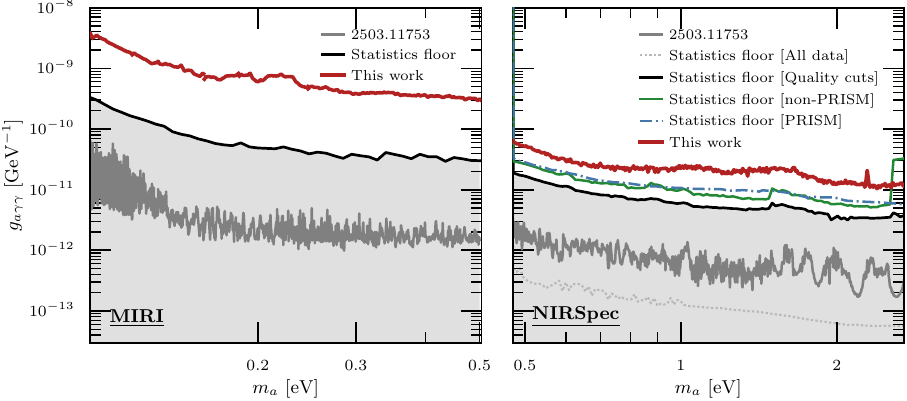}
    \caption{Statistics-floor (Asimov) projections constructed from the dataset released with Ref.~\cite{Pinetti:2025owq}, compared with the constraints published in that work (gray) and with the constraints derived in this work (red), which are based on the substantially larger sample of our primary analysis and are shown for reference. \textbf{Left panel:} Asimov projections and analysis results (smoothed for clarity) for the MIRI instrument. The statistics floor of the complete MIRI dataset of \cite{Pinetti:2025owq} as assessed by Asimov projection is shown in black and lies roughly an order of magnitude above the published constraint at all masses. No analysis of the released data could have attained the published sensitivity. \textbf{Right panel:} Asimov projections and analysis results for the NIRSpec instrument. When every released product and its reported uncertainties from the dataset of \cite{Pinetti:2025owq} are taken at face value (light dotted), the Asimov projected sensitivity appears nominally compatible with the comparatively weaker published constraint. However, the projected sensitivity is driven by the ten pathological products of Table~\ref{tab:nirspec_quality_cuts}; removing only those ten products yields the quality-cut floor (black), which, as in the case of MIRI, lies above the published constraint at every mass. Restricting further to observation-level Stage-3 products, we show the floors of the PRISM (blue dash-dotted) and non-PRISM (green) subsamples separately; the PRISM subsample nominally carries information comparable to all other dispersers combined across much of the mass range. As before, shaded regions below the black curves cannot be reached by any analysis of the released data.}
    \label{fig:statistics_floor}
\end{figure*}

The sensitivities reported in Ref.~\cite{Pinetti:2025owq} can also be compared directly with the statistical information contained in the spectra released with that work. For this purpose, we construct an idealized Asimov projection~\cite{Cowan:2010js} using only this restricted released dataset. This exercise is distinct from our primary analysis, which uses a substantially larger JWST dataset.

We assume that the joint astrophysical and instrumental backgrounds are perfectly known and that the reported spectral uncertainties are independent Gaussian errors. Because the decay flux scales as $\g^2$, the predicted signal in spectral bin $i$ at coupling $\g$ is $(\g/g_{\rm ref})^2 s_i$, where $s_i$ is the signal evaluated at the reference coupling $g_{\rm ref} = 10^{-10}\,\mathrm{GeV}^{-1}$. For an idealized dataset containing no signal, excluding a coupling $\g$ at 95\% confidence requires
\begin{equation}
    \left(\frac{\g}{g_{\rm ref}}\right)^4 \mathcal{I}
    \geq
    \Delta\chi^2_{95},
    \qquad
    \mathcal{I}
    =
    \sum_i
    \frac{s_i^2}{\sigma_i^2},
\end{equation}
where $\mathcal{I}$ is the Fisher information carried by the released spectra, the sum runs over all bins of all released spectra, and $\Delta\chi^2_{95} = 3.84$ matches the limit-setting convention of Ref.~\cite{Pinetti:2025owq}. The projected 95\% CL sensitivity is therefore
\begin{equation}
    g_{a\gamma\gamma}^{\rm floor}
    =
    g_{\rm ref}
    \left(
        \frac{\Delta\chi^2_{95}}{\mathcal{I}}
    \right)^{1/4}.
\end{equation}
We refer to this projection as the ``statistics floor.'' Because the Fisher information sets the minimum variance attainable by any unbiased estimator of the signal normalization through the Cramér-Rao bound, this floor represents the optimal sensitivity that any analysis of the released spectra could achieve under the stated assumptions. It is moreover deliberately optimistic: it neglects uncertainty in the continuum model, correlations between spectral bins, dust attenuation, and other nuisance parameters. In addition, when constructing the all-data projection, we treat every released spectrum as statistically independent, even when Stage-2 and Stage-3 products contain overlapping exposures. These assumptions generally increase the nominal statistical sensitivity.

\subsubsection{MIRI sensitivity floor}
\label{sec:floor_miri}

The statistics floor obtained from the complete released MIRI dataset of \cite{Pinetti:2025owq}, shown in the left panel of Fig.~\ref{fig:statistics_floor}, is weaker than the published constraint throughout the mass range considered, typically by more than an order of magnitude, despite the deliberately optimistic assumptions above. The data volume of the release therefore cannot produce the claimed sensitivity: the discrepancy is information-theoretic and cannot be attributed to loss of sensitivity from continuum marginalization or other nuisance parameters in an analysis. For reference, the same panel shows the constraints derived in this work. Because our fiducial MIRI sample is substantially larger than the released dataset, its statistics floor does not bound our limits. Nevertheless, they lie above it, illustrating the additional gap between an idealized Fisher projection and the sensitivity realizable once continuum modeling, data-driven error rescaling, and the correction for systematic mismodeling are included.

\subsubsection{NIRSpec sensitivity floor}
\label{sec:floor_nirspec}

\begin{table*}[!htb]
\centering
\small
\setlength{\tabcolsep}{5pt}
\begin{ruledtabular}
\begin{tabular}{llcc}
Target & Data product & Stage 2 & Exposure [ks] \\
GCRV 21765 &
\texttt{jw01409-o133\_t011\_nirspec\_prism-clear\_x1d.fits} &
No & 0.13 \\
GCRV 21765 &
\texttt{jw01409-o218\_t011\_nirspec\_prism-clear\_x1d.fits} &
No & 0.52 \\
GCRV 21765 &
\texttt{jw01409018001\_02101\_00001\_nrs1\_x1d.fits} &
Yes & 0.13 \\
GCRV 21765 &
\texttt{jw01409018001\_02101\_00002\_nrs1\_x1d.fits} &
Yes & 0.13 \\
GCRV 21765 &
\texttt{jw01409018001\_02101\_00003\_nrs1\_x1d.fits} &
Yes & 0.13 \\
GCRV 21765 &
\texttt{jw01409133001\_02101\_00001\_nrs1\_x1d.fits} &
Yes & 0.13 \\
GCRV 21765 &
\texttt{jw01409218001\_02101\_00002\_nrs1\_x1d.fits} &
Yes & 0.13 \\
GCRV 21765 &
\texttt{jw01409218001\_02101\_00003\_nrs1\_x1d.fits} &
Yes & 0.13 \\
-- &
\texttt{jw01947-c1008\_t015\_nirspec\_g395m-f290lp\_x1d.fits} &
No & 0.93 \\
-- &
\texttt{jw01947-c1010\_t015\_nirspec\_g395m-f290lp\_x1d.fits} &
No & 0.93 \\
\end{tabular}
\end{ruledtabular}
\caption{
NIRSpec data products excluded by the quality cuts used in constructing the statistics-floor projection for the restricted dataset released with Ref.~\cite{Pinetti:2025owq}. This table therefore refers only to the data products used in our study of that released dataset and not to the larger NIRSpec sample used in our primary analysis. Target names and exposure times are taken directly from the released data archive; no target name was reported for the two \texttt{jw01947} products, which are candidate-type (c-type) association products of the same underlying observation and are identical to one another within the release. ``Stage 2'' indicates whether the listed spectrum is an exposure-level Stage-2 data product rather than a Stage-3 combined product. Including both a Stage-2 product and a Stage-3 product containing the same exposure results in double counting of that exposure, see Sec.~\ref{sec:archival_comparison}. The ten excluded products contain a combined $3.3$~ks of integration time; for comparison, the full released NIRSpec sample totals $3{,}830$~ks of tabulated exposure, of which $1{,}934$~ks belong to the Stage-3 products alone. The quality cuts therefore remove less than $0.1\%$ of the released data volume.}
\label{tab:nirspec_quality_cuts}
\end{table*}

The corresponding NIRSpec exercise requires an additional data-quality qualification. When all 3,639 released NIRSpec products and their quoted uncertainties are used at face value, the resulting projection, shown as the light dotted curve in the right panel of Fig.~\ref{fig:statistics_floor}, appears nominally compatible with the published constraint. However, closer inspection reveals that this projected sensitivity is almost entirely driven by ten datasets with anomalously small reported uncertainties. To determine whether these spectra are simply exceptionally precise observations, we compared their reported uncertainties with the scatter observed in the spectra themselves after removing a smooth local continuum. For eight of the ten flagged products, the observed scatter is typically six or more orders of magnitude larger than the reported error bars, while the remaining NIRSpec spectra show discrepancies of order unity to tens. The final two flagged products contain too few spectral bins for such a comparison; they were instead identified by their reported uncertainties alone, in which isolated bins carry errors many orders of magnitude smaller than the neighboring bins of the same spectrum.

Of these ten datasets, eight originate from a single Cycle~0 commissioning program which repeatedly observed the standard star GCRV 21765 for contamination monitoring during 2022 March and April, months before the start of science operations; these are the earliest observations contained in the release. The remaining two are candidate-type association products of a single science observation which are identical to one another within the release. Notably, none of the ten excluded products enters the sample of our primary analysis: six are exposure-level Stage-2 products, which we exclude in favor of the Stage-3 combinations containing the same exposures; two are PRISM observations, which our fiducial analysis excludes; and two are candidate-type association products, which we exclude in favor of observation-level products. Indeed, these two candidate-type products are no longer available on the MAST archive at all. We therefore additionally show a NIRSpec statistics-floor projection after applying a minimal set of quality cuts that removes these ten pathological NIRSpec datasets, as listed in Table~\ref{tab:nirspec_quality_cuts}. This yields the quality-cut projection shown in black in the right panel of Fig.~\ref{fig:statistics_floor}. The published NIRSpec bound of Ref.~\cite{Pinetti:2025owq} is stronger than this floor at every mass we consider, typically by nearly an order of magnitude. As in the MIRI comparison, no analysis of the surviving data could have attained the published sensitivity.

Finally, in the right panel of Fig.~\ref{fig:statistics_floor}, we further decompose the released information by disperser. For this comparison we restrict the released sample to its 326 observation-level Stage-3 products, removing the 3,277 exposure-level Stage-2 products, the 34 candidate-type association products whose exposures duplicate the observation-level combinations, and the two pathological PRISM products identified above; this restriction parallels the construction of our own analysis sample. Within this restricted sample, the 48 PRISM spectra nominally carry statistical information comparable to that of the 278 spectra of all other dispersers combined across much of the mass range, and greater information at the highest masses, where deep grating coverage ends and the non-PRISM floor deteriorates sharply. At face-value, the PRISM data then might contribute appreciably or even dominantly to the information budget of a full archival analysis; we examine this possibility with a dedicated analysis of the PRISM data in Sec.~\ref{sec:prism_analysis}.

\subsection{PRISM data analysis}
\label{sec:prism_analysis}

\begin{figure*}[!htb]
    \centering
    \includegraphics[width=.99\linewidth]{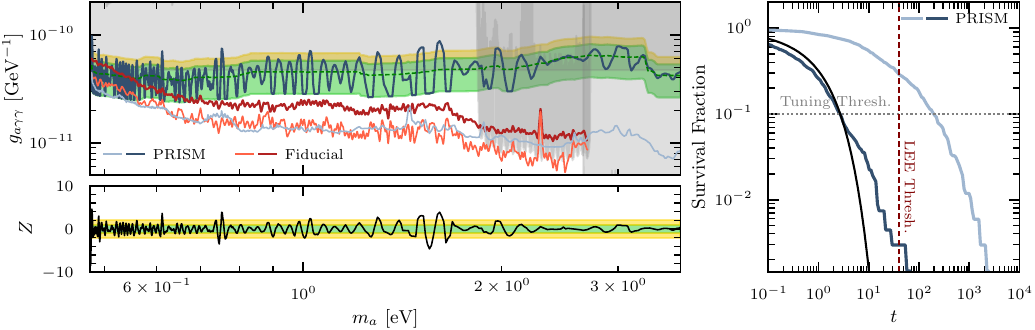}
    \caption{
    Results of our analysis of archival PRISM observations.
    \textbf{Top left:} The 95\% CL upper limit on the axion-photon coupling from the joint PRISM analysis after the spurious-signal correction is shown in dark blue. The green and gold bands denote the expected one- and two-sigma containment intervals of the corrected limit, with the dashed green curve showing the median expectation. The light-blue curve shows the expected limit before the spurious-signal correction and thus the nominal statistical sensitivity of the PRISM sample. For comparison, the corrected limit from the fiducial NIRSpec analysis, smoothed as a function of mass, is shown in dark red, with its corresponding pre-correction expectation in light red. Shaded gray regions denote existing constraints, following Fig.~\ref{fig:money_plot}. \textbf{Bottom left:} Sign-weighted detection significance $Z$ as a function of mass after the spurious-signal correction, with green and gold bands denoting the local one- and two-sigma ranges. \textbf{Right:} Survival functions of the PRISM discovery test statistic $t$, evaluated without setting $t=0$ for negative $\hat{A}_\mathrm{sig}$, before (light blue) and after (dark blue) the spurious-signal correction. The black curve shows the $\chi^2_1$ expectation under the null hypothesis. The dotted gray line marks the $10\%$ survival fraction used to tune the spurious-signal hyperparameter, while the dashed red line marks the LEE-calibrated $5\sigma$ detection threshold. After correction, one mass remains above this threshold, as discussed in the text.
    }
    \label{fig:prism_results}
\end{figure*}

Motivated by the observation that the PRISM data released in Ref.~\cite{Pinetti:2025owq} yielded Asimov projected sensitivities comparable to, or stronger than, those of the accompanying grating data, we also analyze archival PRISM observations. As of August 20, 2026, the archive contained 394 publicly accessible observation-level Stage-3 PRISM products. Of these, 202 had background subtraction applied in the public reduction. Following Sec.~\ref{sec:NIRSpec_Data_Reduction}, we regenerate these products from the Stage 0 data with the background-subtraction steps disabled. Because our likelihood rescales the reported uncertainties in a data-driven manner, the analysis is insensitive to anomalously small quoted error bars.

The resulting constraints are shown in Fig.~\ref{fig:prism_results}. After the spurious-signal correction, the expected 95\% CL limit from the joint PRISM analysis is approximately $4\times10^{-11}\,\mathrm{GeV}^{-1}$ or weaker across the PRISM mass range, and is weaker than the fiducial NIRSpec constraint at nearly all overlapping masses. Before the correction, however, the expected sensitivities of the two analyses are comparable, consistent with the nominal statistical power discussed in Sec.~\ref{sec:statistics_floor}. The substantially weaker corrected PRISM sensitivity therefore reflects a larger degree of systematic mismodeling rather than reduced statistical power.

Quantitatively, the inferred spurious-signal scale $\sigma_\mathrm{nuis}$ reaches up to $60\,\sigma_A$ across the PRISM mass range, compared with factors of $2$--$3$ in the fiducial NIRSpec analysis of Sec.~\ref{sec:data_driven_errors}. The low spectral resolution of PRISM also limits the localization of such mismodeling, since the window required to tune the spurious-signal contribution can become large relative to the wavelength extent of an individual spectrum. Correspondingly, the uncorrected PRISM discovery statistics are grossly inconsistent with the null distribution, extending to $t\sim10^3$. As shown in the right panel of Fig.~\ref{fig:prism_results}, the spurious-signal correction restores broad agreement with the expected null distribution. One excess remains above the LEE-calibrated detection threshold, at $m_a=0.477\,\mathrm{eV}$ with $t\simeq72$. This feature lies near the long-wavelength boundary of the PRISM analysis region and is therefore likely an edge effect. No corresponding excess is present in the fiducial NIRSpec analysis at the same wavelength, and we do not consider it further.

The stronger systematic limitations of the PRISM sample support its exclusion from the fiducial analysis. Moreover, since both the fiducial NIRSpec and PRISM analyses are ultimately systematics limited, we do not combine the two samples: such a joint analysis would not provide a meaningful improvement in robust sensitivity.

\section{Conclusion}
\label{sec:Conclusion}

In this work, we have performed a comprehensive search for the decay of eV-scale axion DM in the public archive of JWST NIRSpec and MIRI IFU spectroscopy. Our analysis combines a custom data reduction designed to preserve a spatially homogeneous decay signal, a signal model that accounts for dust extinction and the Doppler structure of the line, and a nonparametric joint likelihood framework with data-driven error estimation and a correction for systematic mismodeling. The search returns a null result: the only high-significance line-like features that survive our analysis are identified astrophysical lines. We exclude QCD axion DM at masses between $500\,\mathrm{meV}$ and $2.5\,\mathrm{eV}$ and set limits on ALP DM complementary to other astrophysical constraints at masses between $100\,\mathrm{meV}$ and $500\,\mathrm{meV}$. Our NIRSpec constraints are the leading limits on axion-like DM in this mass range and, as recast in App.~\ref{app:ModelIndependent}, on eV-scale DM decaying to photons more generally.

Our results also resolve the outstanding discrepancies among prior JWST axion searches. We find good agreement with the analysis of Ref.~\cite{Saha:2025any} once our data-driven corrections are accounted for. By contrast, the considerably stronger constraints of Ref.~\cite{Pinetti:2025owq} are not reproducible, and we demonstrate that they exceed the sensitivity attainable by any analysis of the datasets from which they were derived.

Finally, we find that the significance of line-like systematic mismodeling grows along with the statistical power of the joint analysis. As a result, the robust sensitivity of the JWST archive is set by the fidelity of its instrumental modeling and data reduction rather than by its accumulated exposure. Improvements in sensitivity to decaying DM in this mass range will therefore likely be driven not by continued data taking but by advances in instrumental characterization and data processing, in particular more faithful estimation of the spectral uncertainties, which the data indicate are underestimated by factors of a few, and modeling of the coherent bin-to-bin structure at the scale of the instrumental resolution.

\section*{Acknowledgments}

{\it 
We thank K.~Boddy, S.~Ellis, D.~Hooper, R.~Laha, O.~Ning, and N.~Rodd for helpful discussions. The Center for Computational Astrophysics at the Flatiron Institute is supported by the Simons Foundation. CD is supported in part by the National Science Foundation grant PHY-2609927. AC is supported by an ERC STG grant (“AstroDarkLS”, grant No. 101117510). This work made use of computing resources provided by the Center for High Throughput Computing at the University of Wisconsin, Madison \cite{https://doi.org/10.21231/gnt1-hw21} and by the Texas Advanced Computing Center (TACC) at The University of Texas at Austin. This research used resources of the National Energy Research Scientific Computing Center (NERSC), a Department of Energy User Facility using NERSC award HEP-ERCAP 0037471. 
}

\appendix

\section{Signal injection tests}
\label{app:SignalInjectionTests}

\begin{figure*}[!htb]
    \centering
    \includegraphics[width=.99\linewidth]{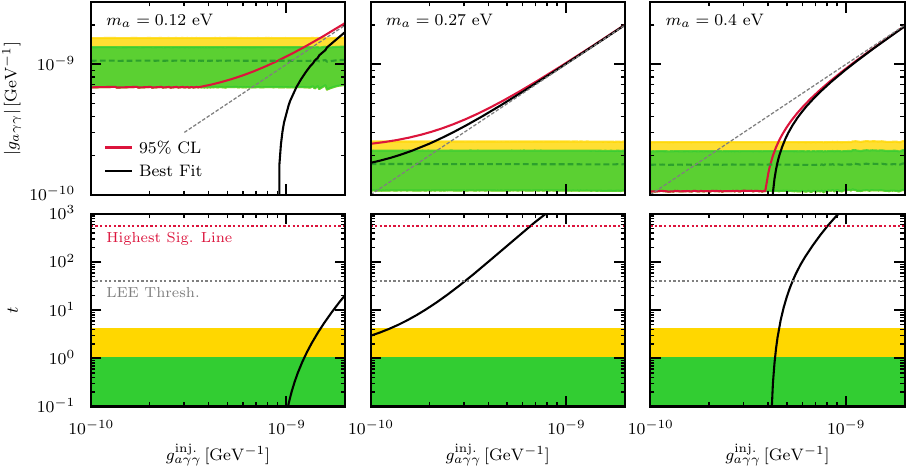}
    \caption{Signal injection tests of the fiducial joint analysis for three representative masses in the MIRI range, $m_a = 0.12$, $0.27$, and $0.4\,\mathrm{eV}$ (left to right). Synthetic signals of varying strength, set by the injected coupling $g^{\rm inj}_{a\gamma\gamma}$, are added atop the real data and the joint likelihood analysis is repeated, subject to the simplifications described in the text; in particular, these tests do not exercise the spurious-signal correction. \textbf{Top panels:} the recovered maximum likelihood estimate of the axion-photon coupling (black) and the recovered 95\% CL upper limit (red) as functions of the injected coupling. We plot the absolute value of the maximum likelihood estimate, which can be driven negative by downward fluctuations of the data. The dashed gray diagonal indicates exact recovery of the injected coupling. Green and gold bands indicate the expected one- and two-sigma containment intervals of the upper limit under the null hypothesis, with the dashed green line indicating the median expected limit. At small injected couplings the recovered limits are consistent with the null expectation, while at large injected couplings the maximum likelihood estimate converges to the injected value. \textbf{Bottom panels:} the discovery test statistic $t$ as a function of the injected coupling. Green and gold bands indicate the ranges corresponding to local significances below one and two sigma, respectively; the dotted gray line indicates the LEE-calibrated threshold for a $5\sigma$ global detection; and the dash-dotted red line indicates the highest local significance at which any of the identified astrophysical lines of Sec.~\ref{sec:Results} appeared prior to the spurious-signal treatment. At sufficiently large injected couplings, the recovered significance exceeds both reference levels. See text for details.}
    \label{fig:miri_signal_injection}
\end{figure*}

\begin{figure*}[!htb]
    \centering
    \includegraphics[width=.99\linewidth]{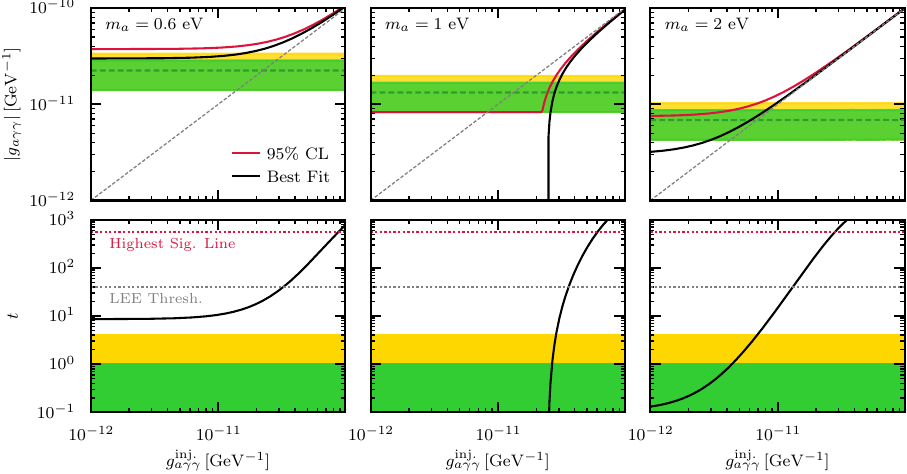}
    \caption{As in Fig.~\ref{fig:miri_signal_injection}, but for a representative set of three masses in the NIRSpec mass range. }
    \label{fig:nirspec_signal_injection}
\end{figure*}

In this Appendix, we perform signal injection tests under our fiducial analysis scheme. For three masses in the MIRI mass range and three masses in the NIRSpec mass range, we inject the expected signal at varying signal strengths atop real data, then apply our analysis procedure to the synthetic data. The results are presented in Fig.~\ref{fig:miri_signal_injection} and Fig.~\ref{fig:nirspec_signal_injection} for MIRI and NIRSpec, respectively. In the top panels, we compare injected signal strength as set by the axion-photon coupling $g_{a\gamma\gamma}$, with the maximum likelihood estimate and 95\% CL upper limit for the axion-photon coupling. Green and gold bands indicate the expected $1\sigma$ and $2\sigma$ containment intervals for the upper limit under the null. In the bottom panels, we provide the value of the discovery test statistic $t$ as a function of injected signal strength, which we compare to the thresholds for $1\sigma$ and $2\sigma$ local significance indicated by the green and gold bands.

In principle, a completely faithful reproduction of our analysis procedure would require that for each injected signal strength, we reanalyze all datasets at all masses in order to self-consistently identify the 25 most constraining datasets for each mass and then re-tune the spurious signal hyperparameter. We instead make two key simplifications, which we subsequently validate, in order to avoid this considerable computational expense.

First, we take the known astrophysical lines of Sec.~\ref{sec:Results}, the strongest of which remain above our detection threshold even after the spurious signal treatment, as evidence that a sufficiently high significance excess will survive the application of this systematic mismodeling correction. We therefore do not consider the role of this procedure as part of our injection tests, and it suffices to analyze only data at the known mass at which we inject the expected signal. Exercising the correction within these tests is possible but computationally expensive, since the large number of masses within the tuning window must be re-analyzed for every injected signal strength. Second, we perform injection and re-analysis on only the 25 datasets which were identified in our fiducial analysis as possessing the greatest sensitivity at the injection mass. In principle, the injection of a signal could change the inferred values of the noise rescaling parameter and GP amplitude in a dataset-dependent way, resulting in an alternate ranking of the most sensitive datasets. However, this is not problematic for the signal injection test for two reasons. Restricting to the previously identified datasets forgoes potentially more sensitive dataset selections. Hence, although our restricted data selection yields a conservative lower bound on the expected sensitivity, the injected signal is discovered at high significance at large signal strengths for all masses considered and depicted in Fig.~\ref{fig:miri_signal_injection} and  Fig.~\ref{fig:nirspec_signal_injection}. Moreover, we show that the expected sensitivity, on which the datasets are ranked, is insensitive to the injected signal strength. Because the rank ordering of datasets by sensitivity is not strongly dependent on the presence of a possible signal, our choice to not re-select datasets for inclusion in the joint analysis is robust.

At small injected couplings, the recovered upper limits are generally consistent with the expectation under the null, while at large injected couplings, the maximum likelihood estimate accurately recovers the injected signal. At a subset of masses, systematic mismodeling produces a high significance deficit, which drives the upper limit to fluctuate downward below the injected coupling at small and moderate signal strengths. In the fiducial analysis, the spurious signal correction is designed to address such spurious high-significance excesses and deficits, with improved coverage of our limits in their presence arising as a byproduct of the increase in the total uncertainty of the analysis; as described above, these injection tests do not exercise that correction, and we make no attempt to demonstrate its role here. At injected strengths large enough to overcome the scale of the systematics, the upper limit no longer excludes the injected coupling, and the discovery test statistic rises rapidly, crossing the LEE-calibrated detection threshold. At even larger couplings, the detection significance safely exceeds the highest significance at which any identified astrophysical line appeared prior to spurious signal treatment.

In total, we take these tests as evidence that our analysis performs at the expected level, and that it would have been capable of detecting any high-significance line-like excesses that may have been present in the data, provided they exceeded the threshold set by the systematic mismodeling we have diagnosed.

\section{Systematic analysis variations}
\label{app:SystematicAnalysisVariations}

In this Appendix, we demonstrate the robustness of our reported limits and detection significances with respect to several choices made in our analysis. In particular, we consider six variations.
\begin{enumerate}
    \item Varying the wavelength ROI from its fiducial value of $\pm 5 \Delta \lambda_\mathrm{sig}$ to $\pm 3 \Delta \lambda_\mathrm{sig}$.
    \item Promoting the parametric background model from a constant background to one linear in the wavelength
    \item Increasing the size of the GP kernel from $\sigma_\mathrm{GP} = 3 \Delta \lambda_\mathrm{sig}$ to  $\sigma_\mathrm{GP} = 4 \Delta \lambda_\mathrm{sig}$
    \item Decreasing the size of the GP kernel from $\sigma_\mathrm{GP} = 3 \Delta \lambda_\mathrm{sig}$ to  $\sigma_\mathrm{GP} = 2 \Delta \lambda_\mathrm{sig}$
    \item Increasing the data subselection from 25 datasets to 50 datasets
    \item Decreasing the data subselection from 25 datasets to 10 datasets
\end{enumerate}
For simplicity, when we perform these analysis variations, we reuse the per-mass ranking of observations by expected sensitivity from our fiducial analysis rather than performing a new end-to-end analysis procedure as in App.~\ref{app:SignalInjectionTests}, constructing each joint analysis from the top 25 datasets in that ranking except in the variations that explicitly modify this number. We iterate over these systematic variations one-by-one such that in our subsequent comparisons, only one aspect of the analysis is changed at a time with respect to our fiducial choices.

The coupling sensitivities derived for each of these analysis variations are compared in the left panel of Fig.~\ref{fig:analysis_bkg_line_variations}. A high degree of consistency is observed, demonstrating the robustness of our sensitivities with respect to the choices made to define our fiducial analysis within the context of this work.

\section{Identification of known astrophysical lines}
\label{app:KnownLines}

A search for a DM decay line must contend with the many astrophysical emission features present in IR sky spectra. To this end, we assemble a rest-frame catalog of known spectral features spanning the combined NIRSpec and MIRI wavelength coverage. The catalog includes the hydrogen recombination series from the Balmer through the Humphreys series and lines of He~I and He~II; forbidden and permitted metal lines; the rovibrational and pure rotational transitions of H$_2$, together with H$_2$O features and the stellar CO bandheads; the mid-IR fine-structure lines of abundant metals; and high-ionization coronal lines. Line wavelengths are drawn from the NIST Atomic Spectra Database~\cite{NIST_ASD}, the CHIANTI atomic database~\cite{Dere:1997chianti, DelZanna:2021chianti}, and the HITRAN molecular database~\cite{Gordon:2022hitran}, supplemented by the JWST MIRI/MRS line identifications toward the Galactic center of Ref.~\cite{Vermot:2025gc}.

Table~\ref{tab:astro_lines} summarizes the astrophysical features identified by our analysis, together with their signed detection significances before and after the spurious-signal correction. Five features appear as strong positive excesses: He~I $1.083\,\mu\mathrm{m}$, [Ne~II] $12.81\,\mu\mathrm{m}$, [Ne~III] $15.56\,\mu\mathrm{m}$, the H$_2$ 0--0 S(1) line at $17.04\,\mu\mathrm{m}$, and [S~III] $18.71\,\mu\mathrm{m}$. Four hydrogen recombination lines, Pa$\beta$, Pa$\alpha$, Br$\delta$, and Br$\gamma$, instead appear as significant flux deficits. The spurious-signal correction substantially reduces all of these significances, as intense astrophysical lines enter the data-driven calibration in the same manner as line-like mismodeling. No other feature produces a comparably significant excess in the joint analysis.

\begin{table}[t]
\centering
\small
\setlength{\tabcolsep}{4pt}
\begin{ruledtabular}
\begin{tabular}{crrr}
Line & Rest Wavelength [$\mu\mathrm{m}$] & $Z$ & $Z$ (spur.) \\
\textbf{He~I} & \textbf{1.08}  & \textbf{23.71} & \textbf{18.90} \\
Pa$\beta$  & 1.28  & -11.39 & -5.10 \\
\textbf{Pa}$\bm{\alpha}$ & \textbf{1.88}  & \textbf{-16.80} & \textbf{-6.37} \\
Br$\delta$ & 1.95  & -14.74 & -5.52 \\
Br$\gamma$ & 2.17  & -14.97 & -4.45 \\
{[Ne II]}  & 12.81 & 25.49 & 5.22 \\
{[Ne III]} & 15.56 & 24.14 & 5.78 \\
$\mathbf{H_2}$     & \textbf{17.04} & \textbf{21.62} & \textbf{8.48} \\
{[S III]} & 18.71 & 14.38 & 5.44 \\
\end{tabular}
\end{ruledtabular}
\caption{Summary of the detected astrophysical lines, including their
identification, rest wavelength, and signed detection significance before and
after correction for spurious signals. Bracketed line identifications denote
forbidden transitions. Bolded rows indicate lines which remain above our
LEE-calibrated detection threshold even after the spurious-signal correction.}
\label{tab:astro_lines}
\end{table}

\section{Impact of astrophysical modeling of DM decay lines}
\label{app:line_model_systematics}

\begin{figure*}[!htb]
    \centering
    \includegraphics[width=.99\linewidth]{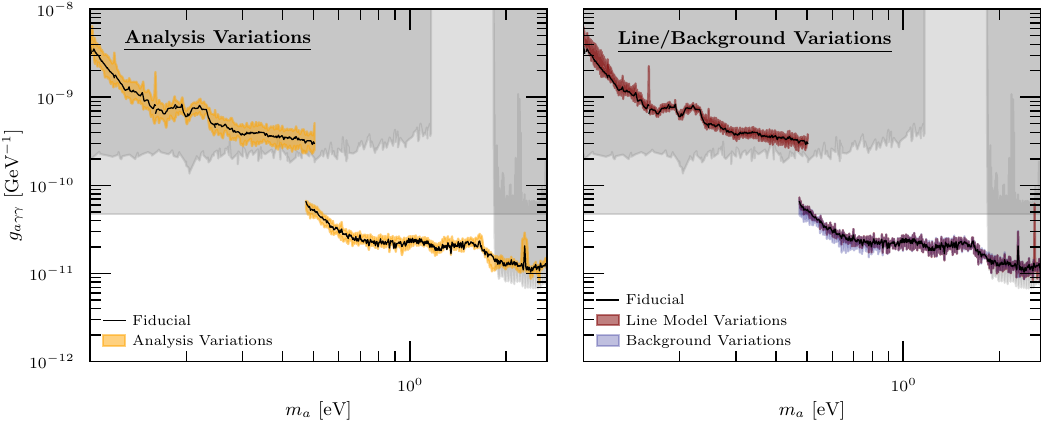}
    \caption{An illustration of the impact of systematic variations of our analysis choices, DM decay line modeling, and data processing on the sensitivity to $g_{a \gamma \gamma}$. Shaded gray regions indicate existing constraints, following the conventions of Fig.~\ref{fig:money_plot}. \textbf{Left panel:} The 95\% CL upper limits from our fiducial MIRI and NIRSpec analyses are shown in black. The gold band spans the full range, from minimum to maximum, of the limits derived at each mass under the six analysis variations described in App.~\ref{app:SystematicAnalysisVariations}, smoothed as a function of mass for visual clarity. The derived limits remain close to the fiducial result under all variations. \textbf{Right panel:} As in the left panel, but for variations to our fiducial line modeling and custom background treatment as part of the data reduction. The red band corresponds to the full range of 95\% CL upper limits set on $g_{a\gamma\gamma}$ as a function of mass under five line modeling choices: the fiducial line model described in Sec.~\ref{sec:IR_Lines_Review} and the systematic variations described in App.~\ref{app:line_model_systematics}. Derived limits differ at only the $\mathcal{O}(10\%)$ level because independent of line modeling choice, the sensitivity is dominated primarily by astrophysically clean off-plane blank-sky observations, where extinction is negligible. The blue band illustrates the full range of 95\% CL upper limits set on $g_{a \gamma \gamma}$ as a function of mass under three choices for NIRSpec data processing: our custom data reduction which disables background subtraction steps described in Sec.~\ref{sec:Data_selection_and_processing} and two variations described in App.~\ref{app:BackgroundSubtraction}. Once again, derived limits differ only at the $\mathcal{O}(10\%)$ level, illustrating the robustness of our results. See text for details.}
    \label{fig:analysis_bkg_line_variations}
\end{figure*}

In this Appendix, we consider systematic variations of our DM decay line modeling. First, we consider two alternatives to our fiducial \texttt{Drimmel03} dust map, holding the wavelength dependence of the extinction fixed to its fiducial power-law form. The \texttt{Combined19} map replaces the fiducial three-dimensional model with the data-driven extinction measurements of Refs.~\cite{2006A&A...453..635M,2019ApJ...887...93G} where they are available, reverting to the \texttt{Drimmel03} model elsewhere, while the \texttt{SFD} map~\cite{1998ApJ...500..525S} instead applies the full integrated column $A_V(\infty)$ uniformly along each line of sight with no distance dependence, maximizing the extinction along each sightline. Both alternatives are provided as part of the \texttt{mwdust} package~\cite{2016ApJ...818..130B}. As a third variation, we retain the fiducial \texttt{Drimmel03} map and consider an alternative treatment of the wavelength dependence, in which the fiducial power law is replaced by the measured $0.09$--$32\,\mu\mathrm{m}$ Milky Way average extinction curve of Ref.~\cite{2023ApJ...950...86G}, which includes the silicate absorption features within the MIRI band. Finally, we also construct limits without any accounting for extinction.

For each of these four systematic variations, we repeat our full analysis over the complete MIRI and NIRSpec datasets, including dataset selection and spurious signal hyperparameter tuning and correction. The span of limits set under these model variations, including those developed under our fiducial line modeling choices, is presented in the red band in the right panel of Fig.~\ref{fig:analysis_bkg_line_variations}. As it must, neglecting dust extinction results in the strongest limits, while the combination of the \texttt{Drimmel03} map with the Milky Way average extinction curve of Ref.~\cite{2023ApJ...950...86G} generally results in the weakest limits. In total, we find that variations in the line model result in only $\mathcal{O}(10\%)$ variations in the constraint set on $g_{a \gamma \gamma}$.

\section{Impact of background subtraction}
\label{app:BackgroundSubtraction}

As described in Sec.~\ref{sec:Data_selection_and_processing}, data-driven background subtractions have the potential to remove a spatially homogeneous DM decay signal from the data, and our fiducial NIRSpec reduction therefore regenerates the 189 Stage-3 spectra whose public reductions applied either the \texttt{bkg\_subtract} or \texttt{master\_background} step with those steps disabled (see Table~\ref{tab:nirspec_background_subtraction}). Aside from the 20 atypical MIRI products noted in Sec.~\ref{sec:MIRI_Data_Reduction}, which we reprocess in the same manner, no background subtraction is applied to MIRI data in the standard processing, so the considerations of this appendix concern NIRSpec alone. We emphasize that the removal of a real DM signal by these subtractions is a possibility we have conjectured from their construction rather than demonstrated; our choice to disable them is precautionary. What we demonstrate in this appendix is that this choice has no meaningful impact on our derived sensitivities.

We consider two variations of our fiducial analysis. In the first, we exclude from the NIRSpec sample the 189 spectra whose public reductions applied a background subtraction, and we re-select the 25 most sensitive datasets at each mass from the remaining spectra, each of which is identical to the corresponding product distributed by the archive. This variation removes any dependence on our custom processing. In the second, we retain the fiducial dataset selection but replace our custom-processed spectra with the default archival products, in which the background subtractions were applied. Because these subtractions could remove a signal, we would not adopt this variation for a search; it serves here only to demonstrate that the background-subtraction settings do not determine the derived sensitivity.

The results of both variations are presented in the blue band in the right panel of Fig.~\ref{fig:analysis_bkg_line_variations}, with minimal differences found between those derived under our fiducial data reduction procedure and the variations considered here. We conclude that the sensitivity of our analysis is not an artifact of our background-subtraction choices: disabling these steps guards against the possible removal of a real signal, while having no meaningful effect on the derived limits.

\section{Model-independent limits on DM decay}
\label{app:ModelIndependent}

\begin{figure}[!htb]
    \centering
    \includegraphics[width=.99\linewidth]{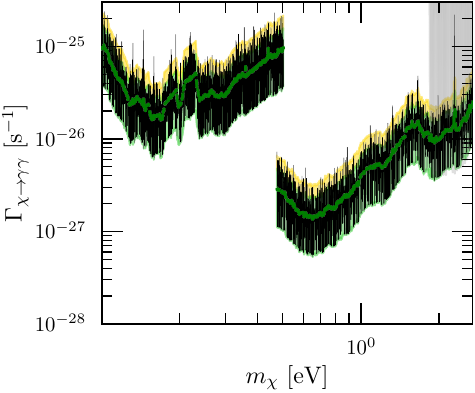}
    \caption{The results of our fiducial analysis translated to constraints on the decay rate to two photons. As before, the black lines correspond to 95\% CL upper limits,  dashed green to the expected 95\% CL upper limit, and the green and gold bands to its 1$\sigma$ and $2\sigma$ containment intervals as a function of mass. WINERED limits at higher masses are also indicated.}
    \label{fig:decay_constraints}
\end{figure}

While in the main text, we have presented our analysis and constraints in terms of the axion-photon coupling $g_{a \gamma\gamma}$, our constraints can be recast as limits on the two-photon decay rate of a generic DM candidate through the relation $\Gamma = \g^2\m^3/64\pi$ of Sec.~\ref{sec:extinction}~\cite{ParticleDataGroup:2026mpi}. We present the constraints developed in our fiducial analysis on this decay rate in Fig.~\ref{fig:decay_constraints}.

\newpage
\bibliography{refs}
\end{document}